\documentclass[dvips,letterpaper,english,prc,reprint,floatfix,showpacs]{revtex4-1}
\usepackage[T1]{fontenc}
\usepackage[utf8x]{inputenc}
\usepackage{color}
\usepackage{verbatim}
\usepackage{textcomp}
\usepackage{mathrsfs}
\usepackage{amsmath}
\usepackage{amssymb}
\usepackage[dvips]{graphicx}

\makeatletter

\providecommand{\tabularnewline}{\\}

 \@ifundefined{textcolor}{}
 {%
   \definecolor{BLACK}{gray}{0}
   \definecolor{WHITE}{gray}{1}
   \definecolor{RED}{rgb}{1,0,0}
   \definecolor{GREEN}{rgb}{0,1,0}
   \definecolor{BLUE}{rgb}{0,0,1}
   \definecolor{CYAN}{cmyk}{1,0,0,0}
   \definecolor{MAGENTA}{cmyk}{0,1,0,0}
   \definecolor{YELLOW}{cmyk}{0,0,1,0}
 }

\renewcommand{\vec}{\mathbf}
\newcommand{\pr}[1]{{\sc{\lowercase{#1}}}}

\makeatother

\usepackage{babel}

\usepackage{hyperref}
\usepackage{dcolumn}

\begin{document}

\title{$\alpha$-decay calculations of heavy nuclei using an effective
Skyrme interaction}

\author{D. E. Ward}\email{daniel.ward@matfys.lth.se}
\author{B. G. Carlsson}
\author{S. Åberg}

\affiliation{Division of Mathematical Physics, LTH, Lund University, P.O.
Box 118, S-22100 Lund, Sweden}
\begin{abstract}
\begin{description}
\item[Background] For nuclei heavier than $^{208}$Pb $\alpha$ decay
is a dominating decay mode, and in the search of new superheavy elements
one often observes chains of $\alpha$ decays.
\item[Purpose] Explore and test microscopic descriptions of $\alpha$ decay
based on theories with effective nuclear interactions.
\item[Methods] The nuclear ground states are calculated with the
Hartree-Fock-Bogoliubov (HFB) method using the Skyrme interaction.
Microscopic $\alpha$-decay formation amplitudes are calculated from
the HFB wave functions, and the $R$-matrix formalism is utilized
to obtain decay probabilities.
\item[Results] Using a large harmonic-oscillator basis we obtain
converged $\alpha$-decay widths. A comparison with experiment including
all spherical even-even $\alpha$ emitting nuclei shows that the model
consistently predicts too small formation amplitudes while relative
values are in good agreement with experiment.
\item[Conclusions] The method was found to be numerically practical
even with a large basis size. The comparison of formation amplitudes
suggests that the pairing type correlations included in the HFB approach
cannot produce sufficient $\alpha$-particle clustering. 
\end{description}
\end{abstract}


\maketitle

\section{Introduction}

Superheavy elements (SHE) can be formed in heavy-ion fusion reactions,
and typically $\alpha$ decay in several steps, see e.g. \cite{Oganessian2007}.
In a recent experiment \cite{RudolphEtAl2013} it has been possible
to measure the emitted $\alpha$ particles in coincidence with $\gamma$ radiation.
This opens up possibilities to identify SHE through x rays, as
well as to obtain detailed spectroscopic information. Such detailed
nuclear structure experiments call for an accurate theoretical description
that simultaneously provides a good prediction of both the structure
of superheavy nuclei and the $\alpha$-decay lifetimes. A good starting
point is then to consider a microscopic model based on interacting
nucleons where both the structure and the reaction parts can be treated
on the same footing. 

Calculations of $\alpha$ decay can be carried out at various levels
of sophistication. Currently, most microscopic approaches are based
on either microscopic-macroscopic models employing Woods-Saxon potentials
combined with BCS pairing, or for some particular nuclei (e.g., $^{212}$Po)
using shell-model approaches where a few valence particles are allowed
to interact via effective model-space interactions, see, e.g., Refs.
\cite{Lovas1998,Delion2010}.

In this work, the structure model is based on modern and well-tested
effective Skyrme interactions which allow for microscopic descriptions
of nuclear properties throughout the nuclear chart. Wave functions
of mother and daughter nuclei are obtained self-consistently using
the Hartree-Fock-Bogoliubov (HFB) method and correlations are modeled
using a density dependent zero-ranged pairing interaction. Taking
the Skyrme interaction as a starting point allows different levels
of correlations, that are particularly important to describe the $\alpha$-particle
formation, to be subsequently included e.g. using the approach of
\cite{Carlsson2013}. 

A microscopic description of the $\alpha$ decay is obtained through
the $R$-matrix approach \cite{Lovas1998,Delion2010,Descouvemont2010}
where the calculated wave functions of the mother and daughter nuclei
are used to project out a formation amplitude for the $\alpha$ particle.
Beyond the range of nuclear forces this amplitude is matched to the
asymptotic Coulomb solution from which the flow of emitted $\alpha$ particles
can be determined.

The method is quite general and can be applied to even-even as well
as to odd nuclei \cite{PoggenburgMangRasmussen1969}. Especially for
odd nuclei it is important to have a reliable microscopic model to
be able to predict the large variations in the half-lives for decays
to different excited states. In this first study, we test the method
for the description of $\alpha$-decaying heavy, spherical even-even
nuclei. 

The paper is organized as follows. In Sec. \ref{sec:Formalism}
the theoretical formalism is described. We give the details of the
nuclear structure model and review and discuss the treatment
of $\alpha$ decay in the $R$-matrix approach. In Sec. \ref{sec:MF-Pairing}
we investigate the convergence of the calculated formation amplitude,
and its dependence on the parameters of the mean field and pairing
force. Calculated $\alpha$ widths are compared to available experimental
data on heavy near-spherical nuclei in Sec. \ref{sec:ComareExp},
where the model also is applied to make predictions for $\alpha$ decay
of the SHE near the predicted shell closures at $N=184$, $Z=114$
and 126. The results are discussed in Sec. \ref{sec:Discussion}
where in particular possible shortcomings and improvements of the
model are considered. Finally, in Sec. \ref{sec:Conclusions}
we conclude and summarize the results.

\section{Formalism\label{sec:Formalism}}

In this section the formalism for our theoretical description of $\alpha$ decay
is discussed. The ingredients of the nuclear structure model are provided
in Sec. \ref{sub:Nuclear-structure-model}. An overview of
the theoretical treatment of the $\alpha$ decay is given in Sec.
\ref{sub:Decay-treatment}, and in Sec. \ref{sub:Formation-amplitude}
we describe how the formation amplitude is obtained.

\subsection{Nuclear structure model \label{sub:Nuclear-structure-model}}

The ground states of the mother and daughter nuclei are described
using the Hartree-Fock-Bogoliubov (HFB) method with an effective Skyrme
interaction in the particle-hole channel \cite{BenderHeenenReinhard2003}.
The HFB equations are solved using an extended version of the program
\textsf{\pr{HOSPHE}} (v1.02) \cite{Carlsson2010p2}. This code works
with a spherical harmonic oscillator basis and can handle large basis
sizes where the maximum oscillator shell included can be as high as
$N_{\mathrm{max}}=70$. A large basis size is essential in order to obtain
convergence for the $\alpha$-particle formation amplitudes.

For the pairing a density-dependent zero-range $\delta$ interaction \cite{Dobaczewski2004}
combined with an energy truncation, $e_{\mathrm{cut}}^{\mathrm{e.s.}}$, in the equivalent spectra
\cite{Stoitsov2005} is adopted. The pairing interaction is parametrized
by

\begin{equation}
V_{\mathrm{pair}}^{q}(\mathbf{r},\mathbf{{r}')}=V_{q}\left[1-\beta\frac{{\rho(\mathbf{\mathbf{{r)}}}}}{\rho_{c}}\right]\delta(\mathbf{{r}-\mathbf{{r}'}}),\; q=n,p,\label{eq:pairingPot}
\end{equation}
where $\rho_{c}=0.16\,\mathrm{fm}^{-3}$ is the saturation density
of nuclear matter and $\beta$ is a parameter determining the density
dependence. In the case of so called surface pairing, i.e., $\beta=1$,
the pairing energy density gets its main contribution from the surface
region. A density independent pairing is obtained when $\beta=0$
in which case the main contribution comes from the nuclear interior.
Pairing is treated both using the HFB approach and with an approximate
version of the Lipkin-Nogami (LN) method \cite{Stoitsov2005}. The
LN method provides an approximate particle-number restoration that
gives more realistic pairing solutions and avoids the collapse of
the pairing for magic nuclei obtained with the HFB method.

The proton pairing strength, $V_{p}$ , is tuned so that the theoretical
odd-even mass difference, $\Delta_{p}^{\mathrm{th}}(N,Z)$, agrees with the
experimental three-point gap centered on the odd nucleus, 

\begin{equation}
\Delta_{p}^{\mathrm{exp}}(N,Z)=E(N,Z+1)-\frac{1}{2}[E(N,Z)+E(N,Z+2)].\label{eq:3ptFormula}
\end{equation}
To have a simple recipe we approximate $\Delta_{p}^{\mathrm{th}}(N,Z)$ by
the lowest quasiparticle energy $E_{p}^{\mathrm{min}}$ calculated for the
even-even nucleus $_{Z}X_{N}$. The same prescription is used for
the neutron pairing strength, $V_{n}$.

\subsection{Decay treatment\label{sub:Decay-treatment}}

$\alpha$ decay is treated microscopically using
the same $R$-matrix-based approach that was used in Refs. \cite{Delion1996,Delion2000,Insolia1988}
and reviewed in Ref. \cite{Lovas1998}. An important feature of $\alpha$ decay
is the tunneling through the long-range Coulomb potential between
the daughter nucleus and $\alpha$ particle. When the $\alpha$ particle
is far away from the daughter nucleus with $Z_{D}$ protons, their
relative motion is described by an outgoing Coulomb wave function,

\begin{equation}
\frac{O_{L}(E,r)}{r}=\frac{1}{r}\left[G_{L}(\eta,\kappa r)+iF_{L}(\eta,\kappa r)\right],\label{eq:OCoulomb}
\end{equation}
where $E$ is the resonance energy, $L$ the angular momentum, $F$ and $G$ the regular and irregular Coulomb
wave functions \cite{AbramowitzStegun}, $\kappa=\frac{\sqrt{2\mu E}}{\hbar}$
and $\eta=\frac{2Z_{D}\mu e^{2}}{\hbar^{2}\kappa}$, where $\mu$ is the reduced mass. In the $R$-matrix
approach the system is divided into inner and outer regions. The solution
for the relative motion in the inner region is matched to this outgoing
Coulomb wave function at a matching radius, $r_{c}$. For the spherical
case, the absolute width, $\Gamma$, of the $\alpha$-decay with energy
$Q_{\alpha}$, is given by
\begin{equation}
\Gamma(r_{c})=2\gamma_{0}^{2}(r_{c})P_{0}(Q_{\alpha},r_{c}),\label{eq:RmatrixGamma}
\end{equation}
where $\gamma_{L}$ is the reduced width, 
\begin{equation}
\gamma_{L}^{2}(r_{c})=\frac{\hbar^{2}}{2\mu r_{c}}r_{c}^{2}g_{L}^{2}(r_{c}),\label{eq:ReducedWidth}
\end{equation}
that depends on the formation amplitude, $g_{L}(r_{c})$. The formation
amplitude describes the relative $\alpha$-daughter motion, and is
obtained from the overlap of the mother nucleus with an $\alpha$-particle
and daughter nucleus separated by the distance $r_{c}$. $P_{L}$
is the Coulomb penetrability factor,

\begin{equation}
P_{L}(Q_{\alpha},r_{c})=\frac{k_{\alpha}r_{c}}{|O_{L}(Q_{\alpha},r_{c})|^{2}},\label{eq:CoulombPeneterabilty}
\end{equation}
where $k_{\alpha}=\frac{\sqrt{2\mu Q_{\alpha}}}{\hbar}$. Both factors
entering Eq. (\ref{eq:RmatrixGamma}) depend on the matching radius,
$r_{c}$. However, in an exact treatment these dependencies cancel
and in the region where nuclear forces can be neglected $\Gamma$
becomes constant. This constant value of the decay width is related
to the half-life, $T_{1/2}$, through the usual
formula $\Gamma=\hbar\:\mathrm{ln}\,2/T_{1/2}.$

The difference compared to earlier works is that
we here obtain the wave functions for the mother and daughter nuclei
entering in the formation amplitude using the Skyrme-HFB model employing
a large harmonic-oscillator basis. To emphasize some of the approximations
in the treatment, we will briefly discuss the main
features of the so called BCS approach to $\alpha$ decay \cite{Lovas1998}.
One can arrive to the formula (\ref{eq:RmatrixGamma}) using either
the Gamow state \cite{Gamow1928,HumbletRosenfeld1961},
or the $R$-matrix formalism \cite{Thomas1954,Descouvemont2010}.
A discussion on the difference between the two approaches, when applied
to proton decay, can be found in \cite{KruppaNazarewicz2004}. The
main steps of the derivation are presented below from a similar perspective
as in \cite{Zeh1963,Mang1964}. 

We describe the mother nucleus, $(M),$ as an exponentially
decaying Gamow state \cite{Gamow1928} 
\begin{equation}
\Psi_{IM}^{(M)}(\xi_{D},\xi_{\alpha},\vec{r}_{\alpha D};t)=\Psi_{IM}^{(M)}(\xi_{D},\xi_{\alpha},\vec{r}_{\alpha D};0)e^{-i(E_{M}-i\frac{\Gamma}{2})t/\hbar},
\end{equation}
where $I$ and $M$ are the spin and spin projection of the mother
nucleus, respectively. The Jacobi coordinate system $\xi_{D},\xi_{\alpha},\vec{r}_{\alpha D}$
corresponds to internal coordinates of the daughter nucleus and the
$\alpha$-particle, and a vector between their centers of mass. $E_{M}$
and $-\Gamma/2$ are the real and imaginary parts of the complex
energy of the Gamow state \cite{Bohm1989}. The state is normalized at $t=0$ within
some finite volume $V$:
\begin{equation}
\int_{V}\left|\Psi_{IM}^{(M)}(\xi_{D},\xi_{\alpha},\vec{r}_{\alpha D};0)\right|^{2}d\xi_{D}d\xi_{\alpha}d\vec{r}_{\alpha D}=1.\label{eq:GamowNorm}
\end{equation}

To find the rate of emitted $\alpha$ particles,
one can start by approximating the mother nucleus as a combined state
of daughter, $(D),$ and valence particles, $(v)$, from which the
$\alpha$ particle is formed (see \hyperref[Appendix:valence]{Appendix}),

\begin{equation}
\Psi_{IM}^{(M)}(t=0)\simeq\mathcal{A}_{Dv}\left\{ \left[\Phi_{J}^{\left(D\right)}(\xi_{D}),\Phi_{L'}^{\left(v\right)}(\xi_{\alpha},\vec{r}_{\alpha D})\right]_{IM}\right\} ,\label{eq:mother}
\end{equation}
where the operator $\mathcal{A}_{Dv}$ \cite{Lovas1998} exchanges
coordinates between the two parts in order to make the state fully
anti-symmetric. 

For large distances between the $\alpha$ particle
and the daughter nucleus, $r_{\alpha D}$, the components of the mother
nucleus that contribute to the $\alpha$-decay width are assumed to
be described by the daughter nucleus wave function, $\Phi_{JM_J}^{(D)}(\xi_{D})$,
the spin zero intrinsic wave function of the $\alpha$ particle, $\Phi_{00}^{(\alpha)}(\xi_{\alpha})$,
and a wave function of their relative motion, $Y_{LM_L}(\hat{r}_{\alpha D})u_{L}(r_{\alpha D})$,
\begin{equation}
\mathcal{A}_{Dv}\left[\Phi_{J}^{\left(D\right)}(\xi_{D}),\Phi_{0}^{\left(\alpha\right)}(\xi_{\alpha})Y_{L}\left(\hat{r}_{\alpha D}\right)u_{L}(r_{\alpha D})\right]_{IM}.\label{eq:channelWaveFunction}
\end{equation}
The formation amplitude, $g_{L}(r_{\alpha D}),$ is defined as the
overlap between the mother nucleus wave function, $\Psi_{IM}^{(M)},$
and the intrinsic and angular parts of expression (\ref{eq:channelWaveFunction}).
With the approximation in Eq. (\ref{eq:mother}),
the formation amplitude can be expressed as

\begin{equation}
\begin{split} & g_{L}(r_{\alpha D})=\sum_{L'}\int\left[\Phi_{J}^{\left(D\right)}(\xi_{D}),\Phi_{0}^{\left(\alpha\right)}(\xi_{\alpha})Y_{L}\left(\hat{r}_{\alpha D}\right)\right]_{IM}^{*}\\
 & \times\left[\Phi_{J}^{\left(D\right)}(\xi_{D}),\Phi_{L'}^{\left(v\right)}(\xi_{\alpha},\vec{r}_{\alpha D})\right]_{IM}d\xi_{D}d\xi_{\alpha}d\hat{r}_{\alpha D}.
\end{split}
\label{eq:FormationAmplitudeIntegral-1}
\end{equation}
In this expression we have neglected the exchange between the $\alpha$
particle and the daughter nucleus. This is a valid approximation if
the orbitals the $\alpha$ particle is expanded in are orthogonal
to the orbitals of the daughter nucleus. Clearly this is not fulfilled
in general but is a good approximation when the $\alpha$ particle
is sufficiently far away from the daughter nucleus. In this work we
furthermore restrict ourselves to decay from ground states in spherical
nuclei where mother and daughter nuclei both have spin zero, $J=0$
and $I=0$. This leads to the simpler form of the formation amplitude,

\begin{equation}
\begin{split}g_{0}(r_{\alpha D}) & =\frac{1}{\sqrt{4\pi}}\int\Phi_{00}^{\left(\alpha\right)*}(\xi_{\alpha})\Phi_{00}^{\left(v\right)}(\xi_{\alpha}\vec{r}_{\alpha D})d\xi_{\alpha}d\hat{r}_{\alpha D}.\end{split}
\label{eq:formamp-2}
\end{equation}

The method of finding the wave functions entering
this expression, further discussed below, involves an expansion in
terms of harmonic oscillator basis functions. This implicitly imposes
boundary conditions that the wave function goes to zero for large
radii which is in principle incorrect. A Gamow state should instead
have outgoing waves as boundary conditions. However, since the $\alpha$
particle has to penetrate a wide and high Coulomb barrier we can assume
\cite{Lovas1998} that the harmonic oscillator basis can provide a
good approximation inside the barrier, and use a matching condition
to impose a tail with the correct asymptotic behavior. 

For large distances, $r_{\alpha D}$, beyond the
range of nuclear forces, the formation amplitude should behave as
an outgoing Coulomb wave, see Eq. (\ref{eq:OCoulomb}),

\textcolor{black}{
\begin{equation}
g_{0}^{\mathrm{ext}}\left(Q_{\alpha}-i\Gamma/2,r_{\alpha D}\right)=C\frac{O_{0}\left(Q_{\alpha}-i\Gamma/2,r_{\alpha D}\right)}{r_{\alpha D}}.\label{eq:g_ext}
\end{equation}
This expression is valid both inside and beyond the Coulomb barrier.
The imaginary part of the energy causes the amplitude of $O_{0}$
to increase with $r_{\alpha D}$. Since $\Gamma$, related to the
decay rate, is very small, this increase of the amplitude may be neglected
inside the barrier. Neglecting the small $\Gamma$, the formation
amplitude, Eq. (}\ref{eq:formamp-2}\textcolor{black}{), is matched to
the external solution, Eq. (}\ref{eq:g_ext}\textcolor{black}{), at the
matching radius $r_{\alpha D}=r_{c}$, giving the total formation
amplitude,}

\textcolor{black}{
\begin{equation}
\begin{split}g_{0}^{\mathrm{tot}}\left(r_{\alpha D}\right) & =g_{0}(r_{\alpha D})\theta\left(r_{c}-r_{\alpha D}\right)\\
 & +g_{0}^{\mathrm{ext}}(Q_{\alpha},r_{\alpha D})\theta\left(r_{\alpha D}-r_{c}\right),
\end{split}
\end{equation}
where $\theta$ are Heaviside functions. The constant $C$ in Eq.
(\ref{eq:g_ext}) is determined by requiring $g_{0}(r_{c})=g_{0}^{\mathrm{ext}}(r_{c})$,}

\textcolor{black}{
\begin{equation}
C=r_{c}\frac{g_{0}(r_{c})}{O_0(Q_{\alpha},r_{c})}.\label{eq:Cmatch}
\end{equation}
From the continuity equation, one can obtain the so-called current
expression \cite{HumbletRosenfeld1961,KruppaNazarewicz2004}. It relates
the width $\Gamma$ to the probability flow, $j_{0}$, at $t=0$ through a surface at $r_{\alpha D}=r_{0}$. Choosing $r_{0}$
to correspond to the volume used for the normalization in Eq. (}\ref{eq:GamowNorm}\textcolor{black}{)
gives }

\textcolor{black}{
\begin{equation}
\begin{split} & \frac{\Gamma}{\hbar}=\frac{i\hbar}{2\mu}r_{0}^{2}\\
 & \times\left[g_{0}^{\mathrm{tot}}\left(r_{0}\right)\frac{\partial g_{0}^{\mathrm{tot}*}\left(r_{0}\right)}{\partial r}-g_{0}^{\mathrm{tot}*}\left(r_{0}\right)\frac{\partial g_{0}^{\mathrm{tot}}\left(r_{0}\right)}{\partial r}\right]\\
 & \equiv j_{0}(r_{0}),
\end{split}
\label{eq:currentExpression}
\end{equation}
where we have assumed that $\alpha$ decay is the only decay channel contributing to the probability flow.
Since we neglect the complex part of the energy the flow through two
different spheres that both enclose the origin is equal, and one may
for simplicity evaluate the flow $j_{0}(r)$ in the $r\rightarrow\infty$
limit. Inserting the asymptotic form of $O_{0}(Q_{\alpha},r)$ for
large $r$ }\cite{AbramowitzStegun}\textcolor{black}{{} in Eq. (}\ref{eq:currentExpression}\textcolor{black}{)
gives}

\textcolor{black}{
\begin{equation}
j_{0}(r_{0})=\lim_{r\rightarrow\infty}j_{0}\left(r\right)=|C|^{2}\frac{\hbar k_{\alpha}}{\mu}.\label{eq:AsymptoticCurrent}
\end{equation}
}Combining Eqs. (\ref{eq:Cmatch}), (\ref{eq:currentExpression})
and (\ref{eq:AsymptoticCurrent}) we recover formula (\ref{eq:RmatrixGamma}),

\textcolor{black}{
\begin{equation}
\Gamma=\frac{r_{c}^{2}g_{0}^{2}(r_{c})\hbar^{2}k_{\alpha}}{|O_{0}(Q_{\alpha},r_{c})|^{2}\mu}=2\gamma_{0}^{2}(r_{c})P_{0}(Q_{\alpha},r_{c}).\label{eq:GammaRmatrix-2}
\end{equation}
}

\subsection{\textcolor{black}{Formation amplitude\label{sub:Formation-amplitude}}}

\textcolor{black}{We use the standard coordinate system, $(\vec{R_{\alpha}},\xi_{\alpha})$,
with $\xi_{\alpha}=(\vec{r}_{\pi},\vec{r}_{\nu},\vec{r}_{\alpha})$,
where \cite{Delion2010}:}

\begin{align*}
 & \vec{r}_{\pi}=\frac{\vec{r}_{1}-\vec{r}_{2}}{\sqrt{2}},\,\,\vec{r}_{\nu}=\frac{\vec{r}_{3}-\vec{r}_{4}}{\sqrt{2}},\\
 & \vec{r}_{\alpha}=\frac{1}{2}\left(\vec{r}_{1}+\vec{r}_{2}-\vec{r}_{3}-\vec{r}_{4}\right),
\end{align*}
\textcolor{black}{and 
\begin{equation}
\vec{R}_{\alpha}=\frac{1}{2}\left(\vec{r}_{1}+\vec{r}_{2}+\vec{r}_{3}+\vec{r}_{4}\right).\label{eq:RalphaCoords-1}
\end{equation}
Here $\vec{r}_{1},\vec{r}_{2}$ are the coordinates for the valence
protons, and $\vec{r}_{3},\vec{r}_{4}$ for the valence neutrons.
The Jacobian for the transformation $(\vec{r_{1}},\vec{r}_{2},\vec{r}_{3},\vec{r}_{4})\rightarrow(\vec{R_{\alpha}},\vec{r}_{\alpha},\vec{r}_{\pi},\vec{r}_{\nu})$
is 1. }

To preserve translational invariance, the valence wave function $\Phi_{00}^{(v)}(\xi_{\alpha},\vec{r}_{\alpha D})$
entering in Eq. (\ref{eq:formamp-2}) should describe the motion of 
the valence particles relative to the daughter.
From the nuclear structure model we obtain shell model type wave functions,
which are localized in a laboratory coordinate system, and we approximate
the formation amplitude, (\ref{eq:formamp-2}), by 

\textcolor{black}{
\begin{equation}
g_{0}(R)=\frac{1}{\sqrt{4\pi}}\int\Phi_{0}^{\left(\alpha\right)*}(\xi_{\alpha})\sqrt{8}\tilde{\Phi}_{00}^{\left(v\right)}(\xi_{\alpha},2\vec{R})d\xi_{\alpha}d\hat{R},\label{eq:formamp-3}
\end{equation}
}where \textcolor{black}{$\vec{R}=\vec{R}_{\alpha}/2$ is the center-of-mass
coordinate of the $\alpha$ particle, and} $\tilde{\Phi}_{00}^{(v)}(\xi_{\alpha},\vec{R}_{\alpha})$
is the valence nucleon wave function of the localized mother nucleus,
discussed in the \hyperref[Appendix:valence]{Appendix}.
The approximation in Eq. (\ref{eq:formamp-3}) consists of making
the substitution $\vec{r}_{\alpha D}\rightarrow\vec{R}$ and using
a localized valence nucleon wave function. This approximation is justified
when the daughter nucleus is heavy relative to the $\alpha$ particle,
and the center of mass parts of the laboratory system wave functions for
mother and daughter nuclei are well localized \cite{Tonozuka1979}.
The factor of $\sqrt{8}$ arises to preserve the normalization of
the valence nucleon wave function, when expressed in the coordinate
$\vec{R}$ \textcolor{black}{\cite{Eichler1965}, as can be seen
from}

\begin{equation}
|\tilde{\Phi}_{00}^{(v)}(\xi_{\alpha},\vec{R}_{\alpha})|^{2}d^{3}R_{\alpha}=|\tilde{\Phi}_{00}^{(v)}(\xi_{\alpha},\vec{R}_{\alpha}(\vec{R}))|^{2}8d^{3}R.
\end{equation}

\textcolor{black}{For the intrinsic $\alpha$-particle wave function
$\Phi_{00}^{\left(\alpha\right)}(\xi_{\alpha})$, we use the standard
approximation \cite{Delion2010},}

\textcolor{black}{
\begin{equation}
\begin{split} & \Phi_{00}^{\left(\alpha\right)}(\mathbf{r}_{\pi},\mathbf{r}_{\nu},\mathbf{r}_{\alpha},s_{1},s_{2},s_{3},s_{4})\\
 & =\left(\frac{1}{b_{\alpha}^{3}\pi^{3/2}}\right)^{3/2}e^{-\frac{r_{\pi}^{2}+r_{\nu}^{2}+r_{\alpha}^{2}}{2b_{\alpha}^{2}}}\\
 & \times[\chi_{\frac{1}{2}}(s_{1}),\chi_{\frac{1}{2}}(s_{2})]_{00}[\chi_{\frac{1}{2}}(s_{3}),\chi_{\frac{1}{2}}(s_{4})]_{00},
\end{split}
\label{eq:alphaWF}
\end{equation}
where $\chi_{\frac{1}{2}}(s)$ are spin wave functions. In order to
agree with electron scattering experiments the oscillator length $b_{\alpha}$
should be chosen as $b_{\alpha}\simeq\sqrt{2}$ fm \cite{Delion2010}
and we adopt the value $b_{\alpha}=1.42$ fm throughout. }

\textcolor{black}{Inserting the approximate valence nucleon wave function,
Eq. }(\ref{eq:ValenceWF}), transformed \textcolor{black}{to relative
and total coordinates \cite{Kamuntavicius2001},} and the \textcolor{black}{$\alpha$-particle
wave function, Eq. }(\ref{eq:alphaWF})\textcolor{black}{{} into Eq.
(}\ref{eq:formamp-3}\textcolor{black}{) gives the final expression
for the formation amplitude 
\begin{equation}
\begin{split}g_{0}(R) & =\frac{1}{\sqrt{2}}\sum_{l_{\pi}j_{\pi}}\sum_{n_{\pi}n_{\pi}'}\frac{X_{n_{\pi}n'_{\pi}}^{l_{\pi}j_{\pi}}\hat{j}_{\pi}^{2}}{\widehat{l}_{\pi}}\sum_{l_{\nu}j_{\nu}}\sum_{n_{\nu}n_{\nu}'}\frac{X_{n_{\nu}n_{\nu}'}^{l_{\nu}j_{\nu}}\hat{j}_{\nu}^{2}}{\widehat{l}_{\nu}}\\
 & \times\sum_{N_{12}n_{12}}\langle N_{12}0,n_{12}0;0|n_{\pi}l_{\pi},n_{\pi}'l_{\pi};0\rangle\\
 & \times\sum_{N_{34}n_{34}}\langle N_{34}0,n_{34}0;0|n_{\nu}l_{\nu},n_{\nu}'l_{\nu};0\rangle\\
 & \times\sum_{N_{\alpha}n_{\alpha}}\langle N_{\alpha}0,n_{\alpha}0;0|N_{12}0,N_{34}0;0\rangle\\
 & \times I_{n_{\alpha}}^{(b,b_{\alpha})}I_{n_{12}}^{(b,b_{\alpha})}I_{n_{34}}^{(b,b_{\alpha})}R_{N_{\alpha}0}^{(b)}(2R),
\end{split}
\end{equation}
where $\hat{j}=\sqrt{2j+1}$ and 
\begin{equation}
\begin{split} & I_{n}^{(b,b_{\alpha})}=\int r^{2}drR_{00}^{(b_{\alpha})*}(r)R_{n0}^{(b)}(r).\end{split}
\end{equation}
$R_{nl}^{(b)}(r)$ is here the radial part of a spherical oscillator
wave function with $n$ nodes and angular momentum $l$, and $b$
denotes the oscillator length used for the basis. }

\section{Dependence on mean field and pairing force\label{sec:MF-Pairing}}

In this section we investigate the dependence of the formation amplitude
on the mean field and pairing force. In Sec. \ref{sub:Convergence-of-formamp}
we check that the dimension of the oscillator basis is sufficient
to obtain correct density at large radii, and that the $\alpha$-particle
formation amplitude converges. The sensitivity of the formation amplitude
to the type of Skyrme force used is studied in Sec. \ref{sub:Skyrme-force-parameters},
the role of approximate particle number correction in the HFB solution
is considered in Sec. \ref{sub:Particle-number-correction},
and the role of surface or volume pairing in Sec. \ref{sub:Density-dependence-of-pairing}.

\subsection{Convergence of the formation amplitude\label{sub:Convergence-of-formamp}}

To have confidence in the numerical results, one must make sure that
the obtained formation amplitude does not depend on the size of the
oscillator basis. The formation amplitude must also be converged for
large separations of $\alpha$ particle and daughter nucleus, so that
nuclear forces between the clusters can be neglected.

This implies several criteria that should be fulfilled for the numerical
calculation, the most obvious being a sufficient accuracy for the
tails of the nuclear wave functions. To satisfy the condition of vanishing
nuclear forces between the clusters, the tails should be accurately
calculated to a distance at least as large as the distance where the
nuclear mean field acting on the valence nucleons becomes negligible.

At the HFB level of approximation the nuclear interactions give rise
to density dependent fields. The local mean fields $V(r)$ \cite{VautherinBrink1972}
for protons and neutrons of double magic lead are shown in 
\begin{figure}
\includegraphics[width=0.9\columnwidth,clip]{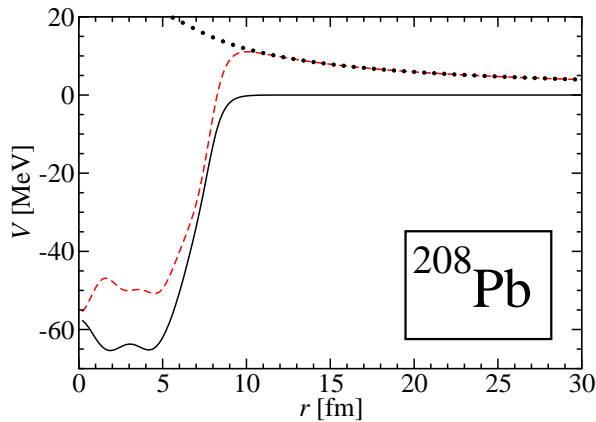}

\caption{(Color online) Local mean fields \cite{VautherinBrink1972} for $^{208}$Pb, from
the SLy4 Skyrme force. The solid line shows the field for neutrons,
the dashed line for protons, and the dotted line shows the Coulomb
part of the proton mean-field. \label{fig:Local-mean-field}}
\end{figure}
 Fig. \ref{fig:Local-mean-field}. They were obtained from a converged
solution of the HFB equations using the code \pr{HFBRAD} \cite{Bennaceur2005}
with the SLy4 Skyrme interaction. The densities from this code are obtained by solving the HFB-equations
in $r$-space in a large box, that give converged results out to
very large radii. It is seen that for $r\geq10$ fm the neutron and
proton nuclear fields are close to zero, and only the Coulomb potential
contributes. Thus the condition of vanishing nuclear forces should
be approximately satisfied at $\alpha$-daughter separations larger
than 10 fm. 

To investigate what size of the spherical oscillator basis is needed
for such wave functions, the neutron and proton densities from using
different number of major oscillator shells are shown in Fig. 
\begin{figure}
\includegraphics[width=0.9\columnwidth,clip]{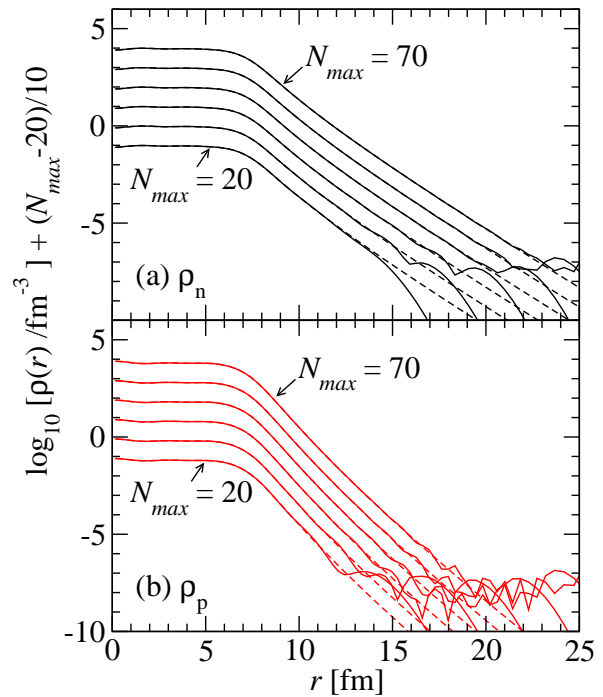}

\caption{(Color online) The upper(lower) panel shows neutron(proton) densities for $^{208}$Pb
obtained by solving the HFB equations using a spherical oscillator
basis (solid lines) and solving on a radial grid (dashed lines). To
separate the different lines, the denisties are multiplied by a factor
$10^{(N_{\mathrm{max}}/10-2)}$. \label{fig:Comparison-of-densitiesN}}
\end{figure}
 \ref{fig:Comparison-of-densitiesN}, where also the results from
\pr{HFBRAD} are shown. Including oscillator shells up to $N_{\mathrm{max}}=20$
gives converged densities out to around 10 fm. It is seen how each
increase of the oscillator size by ten units ($N_{\mathrm{max}}=30,40,\dots$)
increases the convergence radius by an additional 1--2 fm. Similar
trends are found for the pairing density. We find that HFB calculations
for $^{212}$Po give converged pairing density at $r=10$ fm when
$N_{\mathrm{max}}\geq20$. 

The effect of the cutoff in the paring calculation was tested using
cut-off energies $e_{\mathrm{cut}}^{\mathrm{e.s.}}=30,60$ and 90 MeV. When the pairing
strength is tuned so that $\Delta_{\mathrm{exp}}(N=128,Z=84)$, Eq. (\ref{eq:3ptFormula}),
is reproduced, the effect on the formation amplitudes from the different
cutoffs was small, and we shall use $e_{\mathrm{cut}}^{\mathrm{e.s.}}=60$ MeV throughout.

To investigate convergence, the $R$-matrix decay width $\Gamma(r)$
is calculated for $^{212}$Po. The mother and daughter wave functions
are obtained from the SLy4 HFB+LN prescription, and the experimental
$Q_{\alpha}$ value \cite{Akovali1998} is used in the decay width
expression, Eq. (\ref{eq:RmatrixGamma}). As can be seen in Fig. 
\begin{figure}
\includegraphics[width=0.9\columnwidth,clip]{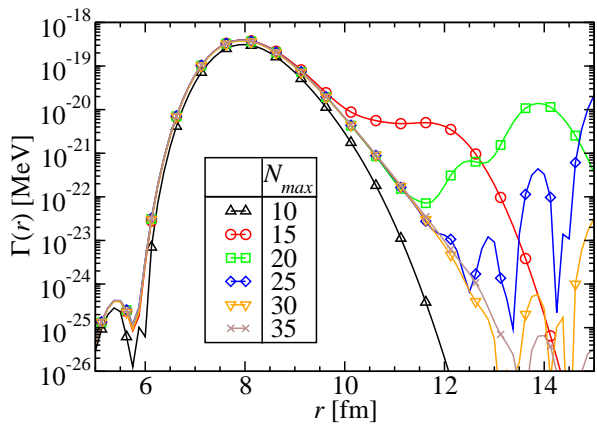}

\caption{(Color online) Decay width $\Gamma$ for $^{212}$Po calculated for different sizes
of the oscillator basis. }

\label{Flo:GammaConvergence} 
\end{figure}
 \ref{Flo:GammaConvergence} the results converge to larger distances
as the basis size is gradually increased from $N_{\mathrm{max}}=10$ to $N_{\mathrm{max}}=35$.
For $N_{\mathrm{max}}=15$ the results are converged to around 9 fm, while
for the largest basis to around 13 fm. By using a basis with $N_{\mathrm{max}}\geq25$
a converged formation amplitude is obtained for separations beyond
the range of inter-cluster nuclear forces. To avoid numerical errors
$N_{\mathrm{max}}=30$ will be used throughout.

\subsection{Skyrme force parameters\label{sub:Skyrme-force-parameters}}

Several fits of Skyrme force parameters exist that give reasonable
results for ground-state observables such as binding energy and rms-radii
\cite{BenderHeenenReinhard2003}. The impact on the microscopic decay
width from the use of different Skyrme forces was tested employing
volume pairing with the LN method. For each Skyrme force the pairing
strength was refitted. The decay width for $^{212}$Po using SLy4,
SKM{*} and SKX interactions are shown in Fig. 
\begin{figure}
\includegraphics[width=0.9\columnwidth,clip]{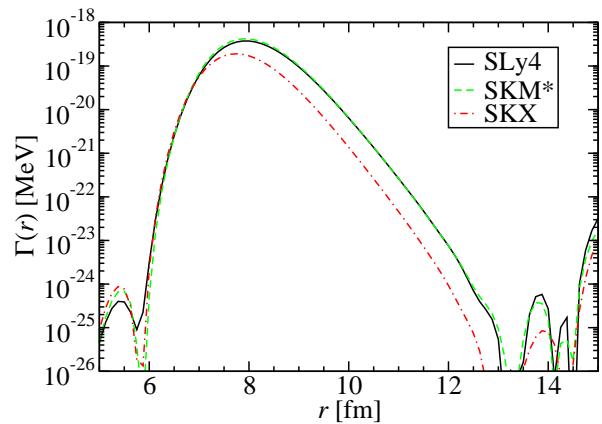}
\caption{(Color online) Decay width $\Gamma$ for $^{212}$Po. Calculated using three different
Skyrme forces. \label{fig:Decay-width-Skyrmes}}
\end{figure}
 \ref{fig:Decay-width-Skyrmes}. The results show a negligible difference
between SLy4 and SKM{*}, while for SKX the decay width is a factor
3.7 smaller at $r=9$ fm. In general, the results are quite insensitive
to the details of the effective particle-hole interaction, and the
SLy4 effective interaction will be used throughout this paper.

\subsection{Particle number correction\label{sub:Particle-number-correction}}

Approximate particle number projection with the LN procedure allows
pairing solutions also when there is a large gap around the Fermi
level in the single-particle spectrum. As discussed below, pairing
correlations have a dramatic effect on the decay widths \cite{Lovas1998}.
Avoiding a collapse of the pairing for magic and semimagic nuclei
the formation amplitude increases considerably. This is illustrated
in Fig. 
\begin{figure}
\includegraphics[width=0.9\columnwidth,clip]{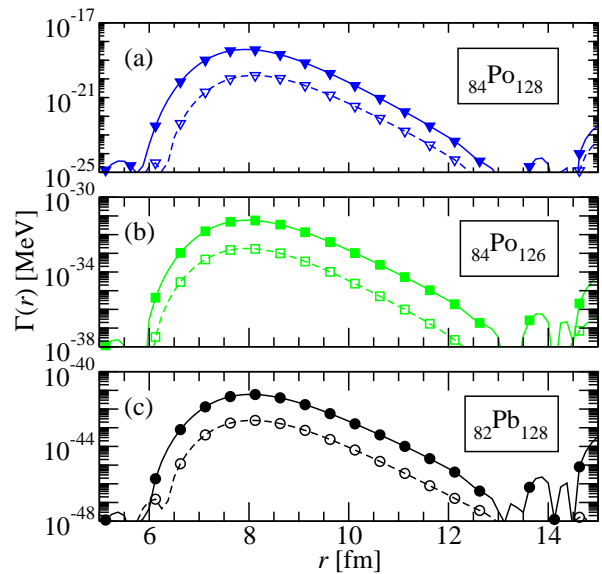}

\caption{(Color online) Effect on decay rate of approximate particle number restoration with
the Lipkin-Nogami procedure. The three panels show results for $^{212}$Po,
$^{210}$Po, and $^{210}$Pb, respectively. The decay width using
HFB+LN wave functions are shown with full lines and solid symbols.
Dashed lines with outlined symbols show results when standard HFB
calculations were performed. \label{fig:LNvsNoLN}}
\end{figure}
 \ref{fig:LNvsNoLN}, where decay widths obtained with and without
the approximate particle number projection are compared. The increase
at $r=9.0$ fm for the g.s. to g.s. $\alpha$ decay of $_{\hphantom{2}84}^{212}$Po$_{128}$,
$_{\hphantom{2}84}^{210}$Po$_{126}$ and $_{\hphantom{2}82}^{210}$Pb$_{128}$,
is a factor 17, 36, and 24, respectively. Two factors influencing
the microscopic decay width are the amount of overlap of the least
bound, or valence, nucleons of the mother nucleus with an $\alpha$ particle,
and the overlap of the remaining nucleons with the daughter. A possible
explanation why the largest enhancement factors are obtained for the
semimagic $^{210}$Po and $^{210}$Pb nuclei is that in these cases
the avoided pairing collapse causes an increase of both types of overlaps
compared to just one type of overlap in the case of $^{212}$Po.

\subsection{Density dependence of pairing force - $^{212}$Po example\label{sub:Density-dependence-of-pairing}}

Since $^{212}$Po has a simple structure with two protons and two
neutrons outside a core of doubly magic lead it is often used to test
microscopic $\alpha$-decay theories. The experimental decay width
for the g.s. to g.s. $\alpha$ decay of this nucleus is $\Gamma_{\mathrm{exp}}=1.53\times10^{-15}$
MeV \cite{Akovali1998}. The converged $R$-matrix decay width shown
in Fig. \ref{Flo:GammaConvergence} is a factor $2.4\times10^{-4}$
smaller than the experimental value at the stationary point around
$r=8$ fm. The down-sloping function $\Gamma(r)$ for larger $r$
also shows that inside the Coulomb barrier the calculated formation
amplitude has a slope corresponding to an $\alpha$ particle that
is considerably more bound to the daughter than observed experimentally. 

Including a density dependence in the effective pairing interaction
allows for a description where the pairing correlations in the surface
of the nucleus is increased, and the correlations in the nuclear interior
is decreased. To see to what extent an increased pairing in the surface
region might favor the formation of $\alpha$ particles, the decay
width of $^{212}$Po was calculated assuming different density dependencies
of the pairing. To get consistent results the pairing strengths are
refitted in each case.

\begin{figure}
\includegraphics[width=0.9\columnwidth,clip]{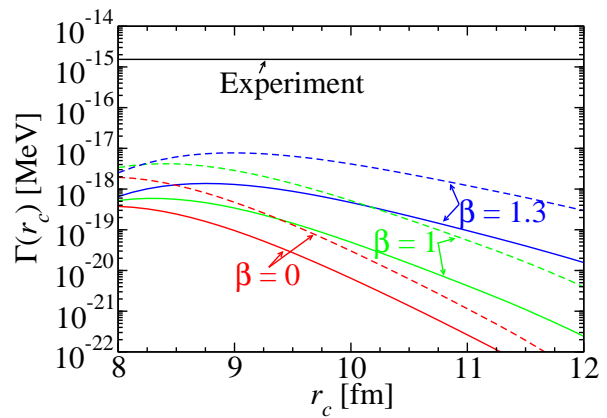}
\caption{(Color online) Decay width for $^{212}$Po.
The effective interaction SLY4 was used together with zero range
pairing with different density dependence: volume $\beta=0$, surface
$\beta=1.0$, anti-volume $\beta=1.3$. The dashed lines show results
for large pairing fit to twice the experimental odd-even gaps. The
straight line shows the experimental value.\label{fig:VaryDensDep}}

\label{fig:VolvsSurfCorrect} 
\end{figure}

The density dependence is determined by the parameter $\beta$ in
Eq. (\ref{eq:pairingPot}), where $\beta=0$ gives volume pairing
and $\beta=1$ amounts to surface pairing. The decay widths obtained
from these two choices are shown in Fig. \ref{fig:VaryDensDep}. One
notices that the width increases by almost one order of magnitude
when surface pairing is used instead of volume pairing. The negative
slope of the decay width is also reduced, indicating that the slope
of the formation amplitude follows the slope of the outgoing Coulomb
wave function slightly better. This corresponds to a formed $\alpha$
particle that is slightly less bound to the daughter nucleus, as compared
to when volume pairing is used. The effect is however not sufficient
to give an $\alpha$-particle amplitude reproducing experimental data. 

Additional clustering in the surface can be introduced by formally
setting $\beta>1$. This corresponds to a force which is repulsive
in the nuclear interior, and strongly attractive in the surface. It
is included as an extreme case; in fact, fits of ground-state properties
suggest that the density dependence of the effective pairing force
should be $0\leq\beta\leq0.5$ \cite{Samyn2003}. Figure \ref{fig:VaryDensDep}
shows that setting $\beta=1.3$ gives an additional increase of the
decay width, as compared to the case of $\beta=1$, although it is
still well below the experimental value.

To test the limits of the pairing force in providing $\alpha$ clustering,
we also show in Fig. \ref{fig:VaryDensDep} (dashed lines) results
of a calculation where the pairing strengths are increased to produce
a gap twice as large as the experimental pairing gap, $\Delta_{\mathrm{th}}=2\Delta_{\mathrm{exp}}$.
It is seen that even in this extreme case the pairing force is unable
to provide a sufficiently large decay width. 

Figure 
\begin{figure}
\includegraphics[width=0.9\columnwidth,clip]{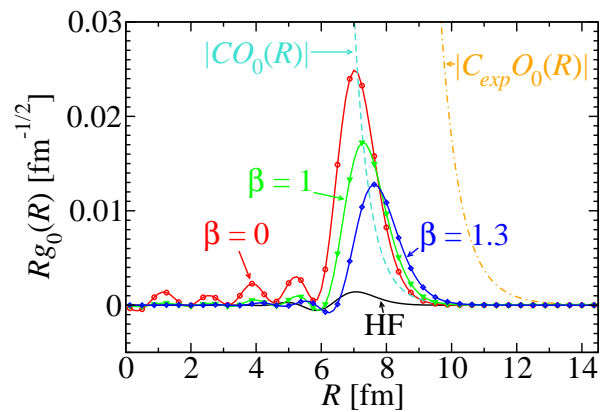}

\caption{(Color online) Formation amplitude $Rg_{0}(R)$ for $^{212}$Po obtained using different
pairing prescriptions. Solid lines show results from microscopic calculations.
Outgoing Coulomb wave functions are shown by dashed and dot-dashed
lines, see text for details.\label{fig:Formation-amplitude-Po212}}
\end{figure}
\ref{fig:Formation-amplitude-Po212} shows the formation amplitudes,
$Rg_{0}(R)$, Eq. (\ref{eq:formamp-3}), for $\beta=0$, 1, and 1.3.
Also shown are results from calculations with negligible pairing,
equivalent to solving the Hartree-Fock equations, denoted by HF. As
can be seen in the figure, the correlations induced by the pairing
greatly increase the formation amplitude compared to the HF results.
At $r=9.0$ fm for the case $\beta=0$ the increase is a factor 16.8,
corresponding to a factor 281 larger decay width.

The modulus of the outgoing Coulomb wave function for the external
region, $Rg_{0}^{\mathrm{ext}}(Q_{\alpha}^{\mathrm{exp}},R)=CO_{0}(R)$, Eq. (\ref{eq:g_ext}),
fitted to the $\beta=0$ formation amplitude at the matching radius
$r_{c}=9$ fm is shown by the dashed line. This outgoing Coulomb wave
function is not a valid solution in the interior of the nucleus, and
increases rapidly with decreasing radius. Examining the tails of the
formation amplitude and the Coulomb wave function, which are too small
to be visible in Fig. \ref{fig:Formation-amplitude-Po212}, we note
that the formation amplitude decreases more rapidly as a function
of $R$, which is the reason why we do not obtain a flat plateau for
$\Gamma(r_{c})$ in Fig. \ref{fig:VaryDensDep}.

Using Eq. (\ref{eq:GammaRmatrix-2})
we can find the external wave function which would perfectly reproduce
experiment. This gives $Rg_{\mathrm{exp}}^{\mathrm{ext}}(Q_{\alpha}^{\mathrm{exp}},R)=C_{\mathrm{exp}}O_{0}(R),$
with $|C_{\mathrm{exp}}|=\sqrt{\frac{\Gamma_{\mathrm{exp}}}{\hbar}\sqrt{\frac{\mu}{2Q_{\alpha}^{\mathrm{exp}}}}}$
, and is shown by the dot-dashed line in Fig. \ref{fig:Formation-amplitude-Po212}.
Comparing the two external wave functions, one notes that to obtain
a plateau for $\Gamma(r_{t})$ with value $\Gamma_{\mathrm{exp}}$ the microscopic
formation amplitudes should be pushed out further beyond the nuclear
surface, and the slope of the tails should be slightly reduced.

\section{Reduced widths compared with experiment\label{sec:ComareExp}}

Even though the model does not produce the right slope and magnitude
of the tail, the formation amplitude depends on the amount of structural
overlap of the mother nucleus with the $\alpha$-daughter configuration.
To be able to reasonably calculate the decay width, some approximate
prescription must be adopted. From the discussion above we see that
the formation amplitude in the nuclear surface must be increased.
Assuming that for all nuclei the correct formation amplitude in the
surface is proportional to the calculated microscopic formation amplitude,
a constant renormalization factor is obtained. Calculated structural
variations in the formation amplitude will then be preserved and the
calculated $\alpha$-decay widths may be compared to experimental
data. Below we perform such an effective description of $\alpha$-decay
widths of all heavy, even-even near-spherical nuclei with measured
decay widths. We also apply the method to predict decay widths for
some $\alpha$-decaying superheavy elements.

The decay widths are calculated using experimental $Q_{\alpha}$ values
from \cite{Akovali1998}. The formation amplitudes are matched to
outgoing Coulomb wave functions in the nuclear surface at the touching
radius $r_{t}$ defined by \cite{Delion1996}

\begin{equation}
r_{t}=r_{0}\left[(A-4)^{1/3}+4^{1/3}\right],\label{eq:TouchingRadius}
\end{equation}
with $r_{0}=1.2$ fm. The touching radius gives an approximate radius
beyond which the $\alpha$ particle and daughter nucleus matter densities
would be separated, which for $^{212}$Po is $r_{t}=9.01$ fm. At
this radius the attractive forces between $\alpha$ and daughter are
not completely negligible (cf. Fig. \ref{fig:Local-mean-field}), but
we find that the normalized decay widths depend weakly on $r_{0}$. 

\begin{figure}
\includegraphics[width=0.9\columnwidth,clip]{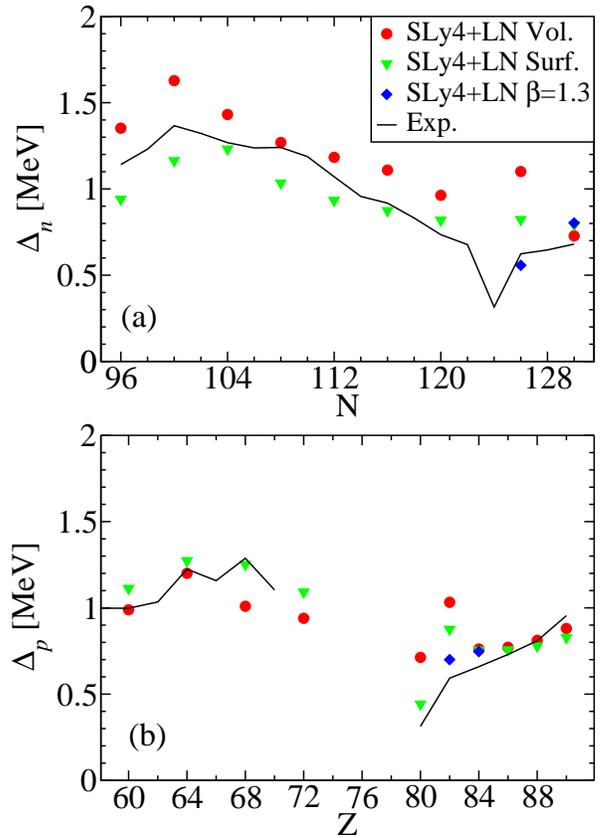}
\caption{(Color online) Upper(lower) panel: neutron(proton) pairing gaps for proton(neutron)-magic
nuclei with neutron(proton) number, $N(Z)$. The theoretical and experimental
pairing gaps, $\Delta_{n(p)}$, are obtained as described in Sec.
\ref{sub:Nuclear-structure-model}. }

\label{fig:SLY4gaps} 
\end{figure}

For the nuclear structure calculation the SLy4 \cite{Chabanat1998}
Skyrme effective nucleon-nucleon potential is used in the particle-hole
channel. The pairing is treated using the Lipkin-Nogami prescription.
Both volume, $\beta=0$, and surface, $\beta=1$, pairing types are
used [Eq. (\ref{eq:pairingPot})]. The pairing strengths used are
shown in Table \ref{tab:Pairing-strengths-1} . Calculated odd-even
gaps for several semimagic nuclei are compared to experiment in Fig.
\ref{fig:SLY4gaps}. 
\begin{table}
\begin{ruledtabular}
\caption{Pairing strengths used in this work. \label{tab:Pairing-strengths-1}}
\begin{tabular}{cccc}
 & $\beta$ & $V_{n}${[}MeVfm$^{3}${]} & $V_{p}${[}MeVfm$^{3}${]}\tabularnewline
\hline 
SLy4  & 0 & -190.5 & -180.5\tabularnewline
SLy4+LN  & 0 & -182 & -175\tabularnewline
SLy4+LN  & 1.0 & -443 & -530\tabularnewline
SLy4+LN & 1.3 & -555 & -770\tabularnewline
\end{tabular}
\end{ruledtabular}
\end{table}
The experimental variation of the pair gap with particle number is
found to be fairly well reproduced by both pairing recipes $\beta=0$
and $\beta=1$. 

Using this prescription the ground state to ground state $\alpha$-decay
widths are determined for all even-even $\alpha$ emitters included
in the compilation of experimental data in \cite{Akovali1998}, and
where the theoretical mass table of M\"{o}ller and Nix \cite{Moller1995}
predicts a near-spherical ground state with quadrupole deformation
parameter $\left|\beta_{2}\right|\leq0.1$ for both mother and daughter
nuclei. This amounts in total to 48 different $\alpha$ emitters. 

\begin{figure}
\includegraphics[width=0.9\columnwidth,clip]{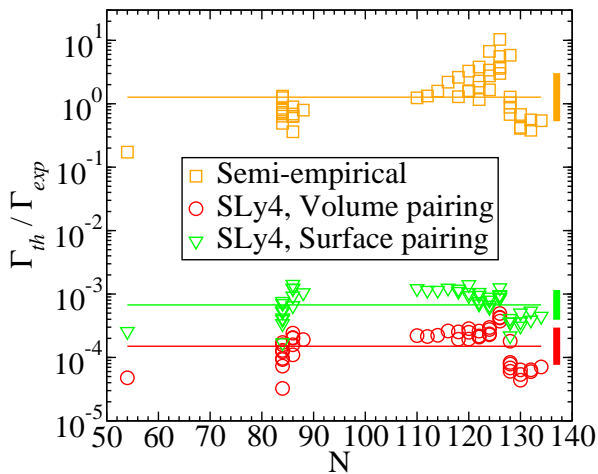}

\caption{(Color online) Theoretical decay widths obtained from the UDL (squares),
SLy4 with volume pairing (circles), and surface pairing (triangles) divided
by the experimental value are shown versus the neutron number of the
decaying nucleus. For each model calculation horizontal lines and
vertical bars denote mean and standard deviation, respectively. \label{fig:GammaThOverGammaExp}}
\end{figure}

The theoretical decay widths, $\Gamma_{\mathrm{th}}=\Gamma(r_{t}),$ divided
by the experimental widths, $\Gamma_{\mathrm{exp}}$, are shown in Fig. \ref{fig:GammaThOverGammaExp}.
For the surface pairing type calculations all 48 near-spherical even-even
$\alpha$ emitters are included, while $^{218}$U is missing from
the volume pairing type calculation due to numerical convergence problems.
For comparison predictions from a semi-empirical model are also shown.
This model, \emph{Universal Decay Law} (UDL) \cite{Qi2009}, is
based on the $R$-matrix expression (\ref{eq:RmatrixGamma}) but the
formation amplitude is parametrized with three free parameters fitted
to data. Here we consider parameter set I, which is fitted to even-even $\alpha$-decay data. 
As seen in Fig. \ref{fig:GammaThOverGammaExp}, for the UDL the results for the ratio $\Gamma_{\mathrm{th}}/\Gamma_{\mathrm{exp}}$
vary around a mean value close to 1. The microscopic models systematically
produce too small decay widths, with slightly better agreement for
the surface pairing. The variation around the mean trend is smaller
for the microscopic models than the results from the UDL, especially
around the $N = 126$ shell closure.

The logarithmic mean deviation, $\mathcal{M}$, from experimental
data,

\begin{equation}
\mathcal{M}=\frac{1}{n}\sum_{i=1}^{n}\mathrm{log_{10}}[\Gamma_{\mathrm{th}}^{(i)}/\Gamma_{\mathrm{exp}}^{(i)}],
\end{equation}
and corresponding standard deviation $\sigma$, are given in Table
\ref{tab:MeanGthovGexp} for each of the calculations.
\begin{table}
\caption{Mean, $\mathcal{M}$, and standard deviation, $\sigma$, of $\mathrm{{log_{10}}}[\Gamma_{\mathrm{th}}/\Gamma_{\mathrm{exp}}]$
for all included nuclei for the three different models in Fig. \ref{fig:GammaThOverGammaExp}:
\label{tab:MeanGthovGexp}}

\begin{ruledtabular}
\begin{tabular}{cdd}
 & \multicolumn{1}{c}{$\mathcal{M}$} & \multicolumn{1}{c}{$\sigma$}\tabularnewline
\hline 
SLy4, Volume pairing & -3.82 & 0.29\tabularnewline
SLy4, Surface pairing & -3.17 & 0.23\tabularnewline
UDL & 0.10 & 0.38\tabularnewline
\end{tabular}\end{ruledtabular}
\end{table}
 The theoretical results using volume or surface pairing underestimates
the decay width by 3 to 4 orders of magnitude, but follow structural
changes in the experimental data quite well. This can be seen by fairly
small $\sigma$ values, that are indeed smaller than those obtained
with the UDL.

The renormalized decay width is now introduced as
\begin{equation}
\tilde{\Gamma}_{\mathrm{th}}^{(i)}=10^{-\mathscr{\mathcal{M}}}\Gamma_{\mathrm{th}}^{(i)},\label{eq:GammaRenorm}
\end{equation}
and correspondingly for the reduced width

\begin{equation}
\tilde{\gamma}^{2}(r_{t})=10^{-\mathscr{\mathcal{M}}}\gamma^{2}(r_{t}).\label{eq:reducedwidth-renormalized}
\end{equation}
In Fig.
\begin{figure}
\includegraphics[clip,width=0.9\columnwidth,clip]{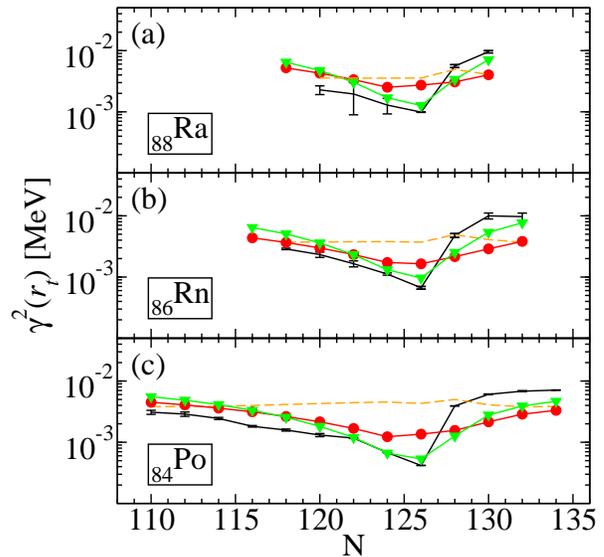}

\caption{(Color online) Reduced width at the touching radius for three isotope chains as a
function of neutron number. The error-bars show extracted experimental
reduced widths. Circles and triangles show the renormalized $\tilde{\gamma}^{2}$,
Eq. (\ref{eq:reducedwidth-renormalized}), obtained from volume and
surface pairing, respectively. The dashed line shows results from
the UDL. \label{fig:Reduced-width-IsotopeChains}}
\end{figure}
\ref{fig:Reduced-width-IsotopeChains} calculated renormalized reduced
widths are compared to experimental data for isotope chains of Po
($Z=84$), Rn ($Z=86$) and Ra ($Z=88$) nuclei. 

The experimental reduced widths show a smoothly decreasing trend as
a function of neutron number towards the shell closure at 126. When
crossing the shell gap, the experimental value increases by about
an order of magnitude, after which there is a smoothly increasing
trend. Comparing volume and surface pairing the smooth behavior in
the open-shell regions is captured fairly well by both pairing models.
However, surface pairing consistently captures the magnitude of the
jump when crossing $N=126$, as well as the trends in the data, better
than the volume type pairing. While the UDL reproduces the
correct mean value, it does not follow the fluctuations around the
shell closure.

Figure
\begin{figure}
\includegraphics[width=0.9\columnwidth,clip]{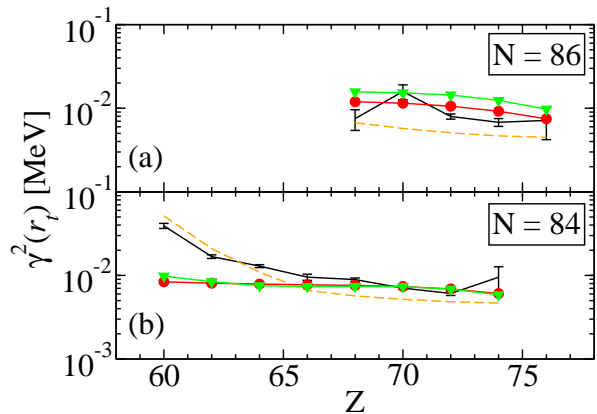}

\caption{(Color online) Similar to Fig. \ref{fig:Reduced-width-IsotopeChains} but for the isobar
chains $N=84$, and $N=86$ as a function of proton number. \label{fig:RedWidthN84}}
\end{figure}
\ref{fig:RedWidthN84} shows the same quantities but for $\alpha$
emitters with neutron numbers 84 and 86. The main deviation from the
experimental trend is that, for the $N = 84$ isotones, the microscopic results fail to capture
the increased formation amplitude with decreasing $Z$. Here the best
agreement with data is obtained from the UDL, suggesting small structural
changes for the $N = 84$ isotones. 

\begin{figure}
\includegraphics[width=0.9\columnwidth,clip]{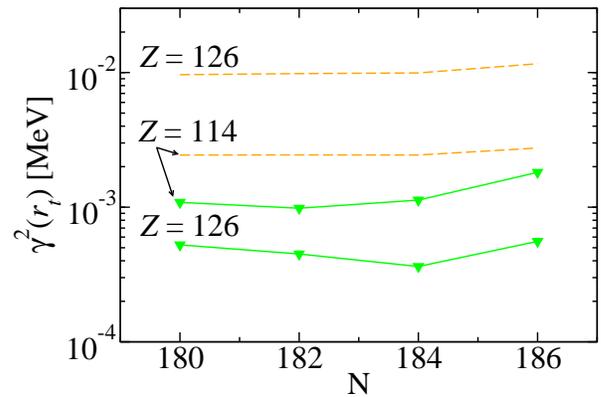}\caption{(Color online) Reduced widths at the touching radius for the SHE of Table \ref{tab:predictionsSHE}.
The triangles show the renormalized microscopic results, from $\beta=1$.
The dashed lines show results obtained using the UDL. \label{fig:RW-SHE}}

\end{figure}

We apply the same prescription to make predictions for the $\alpha$ decay
of the predicted near-spherical superheavy elements with $Z = 114$,
126 and $N = 180$, 182, 184, 186. The microscopic results are obtained
using surface pairing, $\beta=1$, and renormalized using Eqs. (\ref{eq:GammaRenorm})
and (\ref{eq:reducedwidth-renormalized}). The theoretical $Q_{\alpha}$ values
of Refs. \cite{MuntianPatykSobiczewski2003,SobiczewskiPrivateComm}
are used. The predicted half-lives are shown in Table \ref{tab:predictionsSHE}.
\begin{table}
\caption{$T$ denotes half-lives from renormalized microscopic calculations
with $\beta=1$. Predictions from the semi-empirical formula are given
by $T_{\mathrm{UDL}}$.\label{tab:predictionsSHE}}

\begin{ruledtabular}
\begin{tabular}{cdcc}
Nucleus & \multicolumn{1}{c}{$Q_{\alpha}^{th}$ {[}MeV{]}} & $T$ {[}s{]} & $T_{\mathrm{UDL}}$ {[}s{]}\tabularnewline
\hline 
$^{294}114_{180}$ & 9.11 & 593 & 264\tabularnewline
$^{296}114_{182}$ & 9.13 & 523 & 210\tabularnewline
$^{298}114_{184}$ & 9.09 & 571 & 264\tabularnewline
$^{300}114_{186}$ & 10.07 & 0.376 & 0.248\tabularnewline
$^{306}126_{180}$ & 16.23 & 2.20$\times10^{-9}$ & 1.20$\times10^{-10}$\tabularnewline
$^{308}126_{182}$ & 16.25 & 2.25$\times10^{-9}$ & 1.03$\times10^{-10}$\tabularnewline
$^{310}126_{184}$ & 16.25 & 2.61$\times10^{-9}$ & 9.52$\times10^{-11}$\tabularnewline
$^{312}126_{186}$ & 16.64 & 4.77$\times10^{-10}$ & 2.29$\times10^{-11}$\tabularnewline
\end{tabular}\end{ruledtabular}
\end{table}
 The corresponding reduced width amplitudes are shown in Fig. \ref{fig:RW-SHE}.
 For the $Z=114$ isotopes, the difference between the microscopic
and semi-empirical reduced width is less than a factor 3, i.e., the
microscopic model does not predict any dramatic structural effect
that might lead to especially long lifetimes for these superheavy
isotopes. The microscopic reduced widths increase by roughly a factor
2 when crossing the $N=184$ gap, similar to the situation for $N=126$,
shown in Fig. \ref{fig:Reduced-width-IsotopeChains}. The much shorter
predicted half-life for $^{300}114_{186}$ compared to $^{294-298}114_{180-184}$
is due to the $\sim1$ MeV larger predicted $Q_{\alpha}$ value for
this nucleus, see Table \ref{tab:predictionsSHE}. Thus, for predictions
of life-times, uncertainty in the predicted $Q_{\alpha}$ values has
a much larger effect than the difference in the reduced widths obtained
from the semi-empirical and microscopic models. 

Extrapolating further to the region around $^{310}126_{184}$, the
results from the two models start to differ more. The $N = 184$ shell
gap implies a cusp in the microscopically calculated reduced widths,
while a smooth behavior is seen for the semi-empirical UDL. On
the average, the reduced widths from the microscopic model are roughly
a factor 20 smaller than the corresponding values from the UDL.
This gives the order of magnitude longer half-lives obtained from
the microscopic calculation.

\section{Discussion\label{sec:Discussion}}

The calculated decay widths show that the decay rates are systematically
under estimated when the HFB formation amplitudes are used. As with
any process depending on tunneling through a Coulomb barrier, the
asymptotics and thus the flow of particles are extremely sensitive
to the decay energy. For heavy nuclei it is a difficult task to predict
observables such as one-particle separation energies and resonances
with sufficient accuracy for spectroscopy. One can then assume that
for a more complicated process such as $\alpha$-particle formation,
the energy dependent tail of the formation amplitude will never be
described with accuracy comparable to the uncertainties in the experimental
measurement of decay energies, and that this problem will exist to
some degree for even the most sophisticated model. To obtain quantitative
agreement with data, some type of renormalization must be employed.

Here we have adopted the simple procedure consisting of using the
experimental $Q_{\alpha}$ value for the outgoing Coulomb wave function,
combined with renormalizing the decay width by multiplying with one
free parameter. The procedure includes a choice of matching radius,
here chosen as the touching radius, as any mismatch in the slope of
the microscopic formation amplitude compared to the Coulomb wave function
produces an $r_{c}$ dependence of the $R$-matrix decay width. 

Examples of other approaches to renormalize the formation amplitude
can be found in the literature: In Ref. \cite{DelionSandulescu2002},
the properties of the single-particle basis were tuned as an effective
prescription to reproduce the correct absolute value and slope of
the formation amplitude. In Ref. \cite{Betan2012}, in addition to
the $R$-matrix expression for $\Gamma(r_{c})$, Eq. (\ref{eq:RmatrixGamma}),
a reaction theoretical prescription was used. In this prescription
the formation amplitude, corrected for anti-symmetrization in the
nuclear interior, was integrated giving a spectroscopic factor. The
decay was then treated on the one-body level using a local optical
model $\alpha$-daughter potential which can be adjusted to produce
the correct resonance energy. 

In order to get an idea what could be improved in the present $\alpha$-decay
approach we list five additional effects that could be taken into account
and try to estimate their influence:
\begin{enumerate}
\item \emph{Antisymmetrization:} The exchange between the $\alpha$ particle
and the daughter nucleus can only be neglected for large separations
$r$. It is possible to modify the formation amplitude to take exchange
effects into account \cite{Fliessbach1975,FliessbachMang1976}. For
small $r$ this results in a large increase of the formation amplitude,
while for large $r$ the modified formation amplitude reduces to the
$g(r)$ used here. The value of $r$ where $g(r)$ starts to be a good approximation
depends on the daughter and $\alpha$-particle wave functions and
thus varies from case to case. Both types of formation amplitudes
where calculated for $^{212}$Po
in \cite{Tonozuka1979,Betan2012}. In \cite{Tonozuka1979}
the correction amounted to an increase by a factor $\approx2$, and
in \cite{Betan2012} a factor $\approx3$ for the formation amplitude
at $r=9$ fm. Such a correction amounts to an increase by a factor
of 4 or 9 respectively in the decay width. 
\item \emph{Center-of-mass (c.m.) corrections:} In this article and in most
previous studies the formation amplitudes are evaluated with shell-model
wave functions instead of intrinsic states where the c.m. motion is
separated out. The effect of correcting for the c.m. motion was studied
in Ref. \cite{Tonozuka1979}. It is clear from the formulas presented
in this reference that the correction is most important for light
nuclei and will increase the absolute values of the formation amplitudes
as well as stretching the formation amplitudes, it will thus move
their maxima to larger radii in better agreement with experiment. 
\item \emph{Exact Coulomb exchange:} In this work we take the direct part
of the Coulomb interaction into account and treat the exchange part
in a Slater approximation \cite{Giuliani2002}. The asymptotic dependence
of the Coulomb potential for a proton should be $v\left(r\right)=\frac{e^{2}\left(Z-1\right)}{r}$
but becomes $v\left(r\right)=\frac{e^{2}Z}{r}$ with the Slater approximation.
Using exact Coulomb exchange will thus change the slope of the potential
felt by the $\alpha$ particle and make it less bound in the calculations
as well as increasing its magnitude somewhat. However, for a heavy
system such as $^{212}$Po the error in the asymptotic Coulomb potential
is a factor of $\sim1.01$, and we estimate that this will have a
tiny effect on the results.
\item $\alpha$-\emph{particle wave function:} The simple form of the $\alpha$-particle
wave function used here is clearly a convenient approximation. One
could consider more complicated forms obtained e.g. from the same
nuclear structure model as used for the decaying nuclei. Although
the present results are not very sensitive to the oscillator width
taken for the $\alpha$ particle, the effect of having a more realistic
wave function is difficult to estimate.
\item \emph{Correlations:} As discussed in Sec. \ref{sub:Density-dependence-of-pairing}
a substantial increase in the formation amplitude can result from
configuration mixing. For the case of $^{212}$Po, shell-model calculations,
e.g., \cite{Tonozuka1979,Delion2000,Betan2012} show better agreement
for the absolute decay width than the present work. However as far
as we know all these pioneering results have been based on schematic
interactions often directly fitted to the nucleus being studied. A
more systematic investigation of these effects using a globally valid interaction 
would thus be very interesting. 
\end{enumerate}

\section{Conclusions\label{sec:Conclusions}}

We have performed a detailed microscopic calculation of $\alpha$-decay
widths. The mother and daughter nuclei where self-consistently described
applying Skyrme's effective interaction, and the decay widths were
calculated in the $R$-matrix formulation. Our results demonstrate
that it is possible to obtain converged formation amplitudes employing
a large harmonic oscillator basis. In contrast to standard observables
such as masses, radii and excitation energies, these formation amplitudes
probe the amount of cluster components present in the nuclear wave
functions in the surface region. The results give a deeper understanding
of the properties of the wave functions and suggest that a Skyrme-HFB
treatment in combination with the $R$-matrix method is insufficient
in order to predict absolute values of the $\alpha$-decay lifetimes.
Although one should note that there are several extensions to the
formalism which can be envisioned and which seem to go in the right
direction of shortening the too long lifetimes predicted. An improved
description was also obtained by modifying the pairing interaction
to increase the correlations in the nuclear surface. It is however
difficult to determine the physical contents of such a prescription.
In general we found that the pairing force is unable to give sufficient
correlations to provide $\alpha$ particle formation amplitudes agreeing
with data.

It is interesting to see that using a constant factor to renormalize
the results leads to a close agreement with experimental data which
is on par with the results from purely semi-empirical formulas. This
suggests that the missing effects, such as additional correlations
needed to increase the probability of the $\alpha$ particle forming
are in a first approximation proportional to the increase of the formation
amplitudes obtained by including the pairing correlations. More work
is needed in order to improve the model, for example by introducing
more correlations, improvements in the decay formalism and/or treatments
of continuum effects. The results presented here may then serve as
a benchmark to evaluate the impact of such extended theories.
\begin{acknowledgments}
B.G.C. and S.\r{A}. thank the Swedish Research Council (VR) for financial
support.
\end{acknowledgments}
\appendix*

\section{Valence particle wave function\label{Appendix:valence}}

In this appendix we discuss the approximation used for the valence
particles in Sec. \ref{sec:Formalism}. The wave functions presented
below are of shell model type, i.e., localized in the laboratory system, and
thus contain contributions from the total center-of-mass motion of
the nucleons. Such wave functions are written using a tilde, e.g.,
$\tilde{\Psi}(X)$. As there is no proton-neutron mixing, the HFB
wave functions are products of proton and neutron HFB vacua. For
each of the particle species we use the expansion given in Ref. \cite{Lovas1998}
to express the mother nucleus as a function of the daughter. For the
HFB case it becomes 
\begin{align}
\left|M;IM\right\rangle  & \approx\sum_{k<k'}X_{k,k'}c_{k}^{\dagger}c_{k'}^{\dagger}\left|D;jm\right\rangle ,\label{eq:valence_expand}
\end{align}
where proton and neutron indices have been omitted for clarity. The
expansion coefficients are given by the two-particle transfer amplitudes,
\begin{equation}
X_{k,k'}=\langle M|c_{k}^{\dagger}c_{k'}^{\dagger}|D\rangle^{*},\label{eq:symXkk-1-1-1}
\end{equation}
where the $c_{k}^{\dagger}$ operator creates a particle in state
$k$, and $k$ is a short-hand notation for the relevant spherical
single-particle quantum numbers $nljm$. The two-particle transfer
amplitudes are evaluated with the Onishi formula \cite{RingSchuck1980},
\begin{equation}
X_{kk'}=\langle M|c_{k}^{\dagger}c_{k'}^{\dagger}|D\rangle^{*}=\langle M|D\rangle^{*}\kappa_{k,k'}^{DM},
\end{equation}
where the overlap $\langle M|D\rangle$ has an undetermined global
phase, which we set to 1. This phase does not affect the calculated
physical observables. The absolute value of the overlap is given by
\begin{equation}
\left|\langle M|D\rangle\right|=\left|\sqrt{\mathrm{det}\mathbf{U}}\right|.
\end{equation}
The pairing density, $\kappa^{DM}$, is given by 
\begin{equation}
\kappa^{DM*}=-U_{D}^{*}(\mathbf{U}^{T})^{-1}V_{M}^{T},
\end{equation}
where $\mathbf{U}$ is defined in terms of the $U$ and $V$ HFB matrices
\cite{RingSchuck1980} of the mother and daughter states, 
\begin{equation}
\mathbf{U}=U_{D}^{\dagger}U_{M}+V_{D}^{\dagger}V_{M}.
\end{equation}
Due to the spherical symmetry imposed on the HFB solutions the amplitudes
simplify to, 
\begin{equation}
X_{nljm,n'l'j'm'}=\delta_{j,j'}\delta_{l,l'}\delta_{m,-m'}(-1)^{j-m}X_{nn'}^{lj},
\end{equation}
and the approximate neutron or proton part of the mother nucleus wave
function can be written
\begin{equation}
\begin{split}\left|M;00\right\rangle  & \approx\frac{1}{2}\sum_{lj}\sum_{nn'}\hat{j}X_{nn'}^{lj}\\
 & \times\left[c_{nlj}^{\dagger},c_{n'lj}^{\dagger}\right]_{00}\left|D;00\right\rangle ,
\end{split}
\end{equation}
where $I=M=0$, and $\hat{j}=\sqrt{2j+1}$. The corresponding representation in coordinate space becomes
\begin{equation}
\begin{split}\tilde{\Psi}_{00}^{M}\left(X_{Z+2}\right) & \approx\frac{1}{2}\sum_{lj}\sum_{nn'}\hat{j}X_{nn'}^{lj}\\
 & \times\mathcal{A}\left\lbrace\left[\tilde{\phi}_{nlj}\left(\vec{r}_{1}\right),\tilde{\phi}_{n'lj}\left(\vec{r}_{2}\right)\right]_{00}\tilde{\Psi}_{00}^{D}\left(X_{Z}\right)\right\rbrace,
\end{split}
\end{equation}
where $X_Z$ and $X_{Z+2}$ are coordinates for the daughter and mother nucleus, respectively.
The approximate valence particle wave function is thus taken as
\begin{equation}
\tilde{\Phi}^{(v)}(\vec{r}_{1},\vec{r}_{2},\vec{r}_{3},\vec{r}_{4})=\tilde{\Phi}^{\left(v_{\pi}\right)}\left(\vec{r}_{1},\vec{r}_{2}\right)\tilde{\Phi}^{\left(v_{\nu}\right)}\left(\vec{r}_{3},\vec{r}_{4}\right),\label{eq:ValenceWF}
\end{equation}
where
\begin{equation}
\begin{split}\tilde{\Phi}^{\left(v_{q}\right)}\left(\vec{r}_{a},\vec{r}_{b}\right) & =\frac{1}{2}\sum_{lj}\sum_{nn'}\hat{j}X_{nn'}^{q,lj}\\
 & \times\mathcal{A}\left\lbrace\left[\tilde{\phi}_{nlj}\left(\vec{r}_{a}\right),\tilde{\phi}_{n'lj}\left(\vec{r}_{b}\right)\right]_{00}\right\rbrace.
\end{split}
\end{equation}


\begin{thebibliography}{10}%
 \makeatletter
 \providecommand \@ifxundefined [1]{%
  \ifx #1\undefined \expandafter \@firstoftwo
  \else \expandafter \@secondoftwo
 \fi
 }%
 \providecommand \@ifnum [1]{%
  \ifnum #1\expandafter \@firstoftwo
  \else \expandafter \@secondoftwo
 \fi
 }%
 \providecommand \enquote [1]{``#1''}%
 \providecommand \bibnamefont  [1]{#1}%
 \providecommand \bibfnamefont [1]{#1}%
 \providecommand \citenamefont [1]{#1}%
 \providecommand\href[0]{\@sanitize\@href}%
 \providecommand\@href[1]{\endgroup\@@startlink{#1}\endgroup\@@href}%
 \providecommand\@@href[1]{#1\@@endlink}%
 \providecommand \@sanitize [0]{\begingroup\catcode`\&12\catcode`\#12\relax}%
 \@ifxundefined \pdfoutput {\@firstoftwo}{%
  \@ifnum{\z@=\pdfoutput}{\@firstoftwo}{\@secondoftwo}%
 }{%
  \providecommand\@@startlink[1]{\leavevmode}%
  \providecommand\@@endlink[0]{}%
 }{%
  \providecommand\@@startlink[1]{%
   \leavevmode
   \pdfstartlink
    attr{/Border[0 0 1 ]/H/I/C[0 1 1]}%
    user{/Subtype/Link/A<</Type/Action/S/URI/URI(#1)>>}%
   \relax
  }%
  \providecommand\@@endlink[0]{\pdfendlink}%
 }%
 \providecommand \url  [0]{\begingroup\@sanitize \@url }%
 \providecommand \@url [1]{\endgroup\@href {#1}{\urlprefix}}%
 \providecommand \urlprefix [0]{URL }%
 \providecommand \Eprint[0]{\href }%
 \@ifxundefined \urlstyle {%
   \providecommand \doi [1]{doi:\discretionary{}{}{}#1}%
 }{%
   \providecommand \doi [0]{doi:\discretionary{}{}{}\begingroup
   \urlstyle{rm}\Url }%
 }%
 \providecommand \doibase [0]{http://dx.doi.org/}%
 \providecommand \Doi[1]{\href{\doibase#1}}%
 \providecommand \bibAnnote [3]{%
   \BibitemShut{#1}%
   \begin{quotation}\noindent
     \textsc{Key:}\ #2\\\textsc{Annotation:}\ #3%
   \end{quotation}%
 }%
 \providecommand \bibAnnoteFile [2]{%
   \IfFileExists{#2}{\bibAnnote {#1} {#2} {\input{#2}}}{}%
 }%
 \providecommand \typeout [0]{\immediate \write \m@ne }%
 \providecommand \selectlanguage [0]{\@gobble}%
 \providecommand \bibinfo [0]{\@secondoftwo}%
 \providecommand \bibfield [0]{\@secondoftwo}%
 \providecommand \translation [1]{[#1]}%
 \providecommand \BibitemOpen[0]{}%
 \providecommand \bibitemStop [0]{}%
 \providecommand \bibitemNoStop [0]{.\EOS\space}%
 \providecommand \EOS [0]{\spacefactor3000\relax}%
 \providecommand \BibitemShut [1]{\csname bibitem#1\endcsname}%
 \bibitem{Oganessian2007}%
   \BibitemOpen
   \bibfield{author}{%
   \bibinfo {author} {\bibfnamefont{Y.}~\bibnamefont{Oganessian}},\ }%
   \bibfield{journal}{%
   \Doi{10.1088/0954-3899/34/4/R01}{\bibinfo {journal} {J. Phys G}}\ }%
   \textbf{\bibinfo {volume} {34}},\ \bibinfo {pages} {R165} (\bibinfo {year}
   {2007})%
   \bibAnnoteFile{NoStop}{Oganessian2007}%
 \bibitem{RudolphEtAl2013}%
   \BibitemOpen
   \bibfield{author}{%
   \bibinfo {author} {\bibfnamefont{D.}~\bibnamefont{Rudolph}}, \bibinfo
   {author} {\bibfnamefont{U.}~\bibnamefont{Forsberg}}, \bibinfo {author}
   {\bibfnamefont{P.}~\bibnamefont{Golubev}}, \bibinfo {author}
   {\bibfnamefont{L.~G.}\ \bibnamefont{Sarmiento}}, \bibinfo {author}
   {\bibfnamefont{A.}~\bibnamefont{Yakushev}}, \bibinfo {author}
   {\bibfnamefont{L.-L.}\ \bibnamefont{Andersson}}, \bibinfo {author}
   {\bibfnamefont{A.}~\bibnamefont{Di~Nitto}}, \bibinfo {author}
   {\bibfnamefont{C.~E.}\ \bibnamefont{D\"ullmann}}, \bibinfo {author}
   {\bibfnamefont{J.~M.}\ \bibnamefont{Gates}}, \bibinfo {author}
   {\bibfnamefont{K.~E.}\ \bibnamefont{Gregorich}}, \bibinfo {author}
   {\bibfnamefont{C.~J.}\ \bibnamefont{Gross}}, \bibinfo {author}
   {\bibfnamefont{F.~P.}\ \bibnamefont{He\ss{}berger}}, \bibinfo {author}
   {\bibfnamefont{R.-D.}\ \bibnamefont{Herzberg}}, \bibinfo {author}
   {\bibfnamefont{J.}~\bibnamefont{Khuyagbaatar}}, \bibinfo {author}
   {\bibfnamefont{J.~V.}\ \bibnamefont{Kratz}}, \bibinfo {author}
   {\bibfnamefont{K.}~\bibnamefont{Rykaczewski}}, \bibinfo {author}
   {\bibfnamefont{M.}~\bibnamefont{Sch\"adel}}, \bibinfo {author}
   {\bibfnamefont{S.}~\bibnamefont{\AA{}berg}}, \bibinfo {author}
   {\bibfnamefont{D.}~\bibnamefont{Ackermann}}, \bibinfo {author}
   {\bibfnamefont{M.}~\bibnamefont{Block}}, \bibinfo {author}
   {\bibfnamefont{H.}~\bibnamefont{Brand}}, \bibinfo {author}
   {\bibfnamefont{B.~G.}\ \bibnamefont{Carlsson}}, \bibinfo {author}
   {\bibfnamefont{D.}~\bibnamefont{Cox}}, \bibinfo {author}
   {\bibfnamefont{X.}~\bibnamefont{Derkx}}, \bibinfo {author}
   {\bibfnamefont{K.}~\bibnamefont{Eberhardt}}, \bibinfo {author}
   {\bibfnamefont{J.}~\bibnamefont{Even}}, \bibinfo {author}
   {\bibfnamefont{C.}~\bibnamefont{Fahlander}}, \bibinfo {author}
   {\bibfnamefont{J.}~\bibnamefont{Gerl}}, \bibinfo {author}
   {\bibfnamefont{E.}~\bibnamefont{J\"ager}}, \bibinfo {author}
   {\bibfnamefont{B.}~\bibnamefont{Kindler}}, \bibinfo {author}
   {\bibfnamefont{J.}~\bibnamefont{Krier}}, \bibinfo {author}
   {\bibfnamefont{I.}~\bibnamefont{Kojouharov}}, \bibinfo {author}
   {\bibfnamefont{N.}~\bibnamefont{Kurz}}, \bibinfo {author}
   {\bibfnamefont{B.}~\bibnamefont{Lommel}}, \bibinfo {author}
   {\bibfnamefont{A.}~\bibnamefont{Mistry}}, \bibinfo {author}
   {\bibfnamefont{C.}~\bibnamefont{Mokry}}, \bibinfo {author}
   {\bibfnamefont{H.}~\bibnamefont{Nitsche}}, \bibinfo {author}
   {\bibfnamefont{J.~P.}\ \bibnamefont{Omtvedt}}, \bibinfo {author}
   {\bibfnamefont{P.}~\bibnamefont{Papadakis}}, \bibinfo {author}
   {\bibfnamefont{I.}~\bibnamefont{Ragnarsson}}, \bibinfo {author}
   {\bibfnamefont{J.}~\bibnamefont{Runke}}, \bibinfo {author}
   {\bibfnamefont{H.}~\bibnamefont{Schaffner}}, \bibinfo {author}
   {\bibfnamefont{B.}~\bibnamefont{Schausten}}, \bibinfo {author}
   {\bibfnamefont{P.}~\bibnamefont{Th\"orle-Pospiech}}, \bibinfo {author}
   {\bibfnamefont{T.}~\bibnamefont{Torres}}, \bibinfo {author}
   {\bibfnamefont{T.}~\bibnamefont{Traut}}, \bibinfo {author}
   {\bibfnamefont{N.}~\bibnamefont{Trautmann}}, \bibinfo {author}
   {\bibfnamefont{A.}~\bibnamefont{T\"urler}}, \bibinfo {author}
   {\bibfnamefont{A.}~\bibnamefont{Ward}}, \bibinfo {author}
   {\bibfnamefont{D.~E.}\ \bibnamefont{Ward}},\ and\ \bibinfo {author}
   {\bibfnamefont{N.}~\bibnamefont{Wiehl}},\ }%
   \bibfield{journal}{%
   \Doi{10.1103/PhysRevLett.111.112502}{\bibinfo {journal} {Phys. Rev. Lett.}}\
   }%
   \textbf{\bibinfo {volume} {111}},\ \bibinfo {pages} {112502} (\bibinfo {year}
   {2013})%
   \bibAnnoteFile{NoStop}{RudolphEtAl2013}%
 \bibitem{Lovas1998}%
   \BibitemOpen
   \bibfield{author}{%
   \bibinfo {author} {\bibfnamefont{R.}~\bibnamefont{Lovas}}, \bibinfo {author}
   {\bibfnamefont{R.}~\bibnamefont{Liotta}}, \bibinfo {author}
   {\bibfnamefont{A.}~\bibnamefont{Insolia}}, \bibinfo {author}
   {\bibfnamefont{K.}~\bibnamefont{Varga}},\ and\ \bibinfo {author}
   {\bibfnamefont{D.}~\bibnamefont{Delion}},\ }%
   \bibfield{journal}{%
   \Doi{10.1016/S0370-1573(97)00049-5}{\bibinfo {journal} {Phys. Rep.}}\ }%
   \textbf{\bibinfo {volume} {294}},\ \bibinfo {pages} {265 } (\bibinfo {year}
   {1998})%
   \bibAnnoteFile{NoStop}{Lovas1998}%
 \bibitem{Delion2010}%
   \BibitemOpen
   \bibfield{author}{%
   \bibinfo {author} {\bibfnamefont{D.~S.}\ \bibnamefont{Delion}},\ }%
   \Doi{10.1007/978-3-642-14406-6}{\emph{\bibinfo {title} {Theory of particle
   and cluster emission}}},\ \bibinfo {series} {Lecture notes in physics}\ No.\
   \bibinfo {number} {819}\ (\bibinfo {publisher} {Springer},\ \bibinfo {year}
   {2010})%
   \bibAnnoteFile{NoStop}{Delion2010}%
 \bibitem{Carlsson2013}%
   \BibitemOpen
   \bibfield{author}{%
   \bibinfo {author} {\bibfnamefont{B.~G.}\ \bibnamefont{Carlsson}}, \bibinfo
   {author} {\bibfnamefont{J.}~\bibnamefont{Toivanen}},\ and\ \bibinfo {author}
   {\bibfnamefont{U.}~\bibnamefont{von Barth}},\ }%
   \bibfield{journal}{%
   \Doi{10.1103/PhysRevC.87.054303}{\bibinfo {journal} {Phys. Rev. C}}\ }%
   \textbf{\bibinfo {volume} {87}},\ \bibinfo {pages} {054303} (\bibinfo {year}
   {2013})%
   \bibAnnoteFile{NoStop}{Carlsson2013}%
 \bibitem{Descouvemont2010}%
   \BibitemOpen
   \bibfield{author}{%
   \bibinfo {author} {\bibfnamefont{P.}~\bibnamefont{Descouvemont}}\ and\
   \bibinfo {author} {\bibfnamefont{D.}~\bibnamefont{Baye}},\ }%
   \bibfield{journal}{%
   \Doi{10.1088/0034-4885/73/3/036301}{\bibinfo {journal} {Rep. Prog. Phys.}}\
   }%
   \textbf{\bibinfo {volume} {73}},\ \bibinfo {pages} {036301} (\bibinfo {year}
   {2010})%
   \bibAnnoteFile{NoStop}{Descouvemont2010}%
 \bibitem{PoggenburgMangRasmussen1969}%
   \BibitemOpen
   \bibfield{author}{%
   \bibinfo {author} {\bibfnamefont{J.~K.}\ \bibnamefont{Poggenburg}}, \bibinfo
   {author} {\bibfnamefont{H.~J.}\ \bibnamefont{Mang}},\ and\ \bibinfo {author}
   {\bibfnamefont{J.~O.}\ \bibnamefont{Rasmussen}},\ }%
   \bibfield{journal}{%
   \Doi{10.1103/PhysRev.181.1697}{\bibinfo {journal} {Phys. Rev.}}\ }%
   \textbf{\bibinfo {volume} {181}},\ \bibinfo {pages} {1697} (\bibinfo {year}
   {1969})%
   \bibAnnoteFile{NoStop}{PoggenburgMangRasmussen1969}%
 \bibitem{BenderHeenenReinhard2003}%
   \BibitemOpen
   \bibfield{author}{%
   \bibinfo {author} {\bibfnamefont{M.}~\bibnamefont{Bender}}, \bibinfo {author}
   {\bibfnamefont{P.-H.}\ \bibnamefont{Heenen}},\ and\ \bibinfo {author}
   {\bibfnamefont{P.-G.}\ \bibnamefont{Reinhard}},\ }%
   \bibfield{journal}{%
   \Doi{10.1103/RevModPhys.75.121}{\bibinfo {journal} {Rev. Mod. Phys.}}\ }%
   \textbf{\bibinfo {volume} {75}},\ \bibinfo {pages} {121} (\bibinfo {year}
   {2003})%
   \bibAnnoteFile{NoStop}{BenderHeenenReinhard2003}%
 \bibitem{Carlsson2010p2}%
   \BibitemOpen
   \bibfield{author}{%
   \bibinfo {author} {\bibfnamefont{B.~G.}\ \bibnamefont{Carlsson}}, \bibinfo
   {author} {\bibfnamefont{J.}~\bibnamefont{Dobaczewski}}, \bibinfo {author}
   {\bibfnamefont{J.}~\bibnamefont{Toivanen}},\ and\ \bibinfo {author}
   {\bibfnamefont{P.}~\bibnamefont{Vesel{\'y}}},\ }%
   \bibfield{journal}{%
   \Doi{10.1016/j.cpc.2010.05.022}{\bibinfo {journal} {Comp.\ Phys.\ Commun.}}\
   }%
   \textbf{\bibinfo {volume} {181}},\ \bibinfo {pages} {1641} (\bibinfo {year}
   {2010})%
   \bibAnnoteFile{NoStop}{Carlsson2010p2}%
 \bibitem{Dobaczewski2004}%
   \BibitemOpen
   \bibfield{author}{%
   \bibinfo {author} {\bibnamefont{J.Dobaczewski}}\ and\ \bibinfo {author}
   {\bibnamefont{P.Olbratowski}},\ }%
   \bibfield{journal}{%
   \Doi{http://dx.doi.org/10.1016/j.cpc.2004.02.003}{\bibinfo {journal} {Comp.\
   Phys.\ Commun.}}\ }%
   \textbf{\bibinfo {volume} {158}},\ \bibinfo {pages} {158} (\bibinfo {year}
   {2004})%
   \bibAnnoteFile{NoStop}{Dobaczewski2004}%
 \bibitem{Stoitsov2005}%
   \BibitemOpen
   \bibfield{author}{%
   \bibinfo {author} {\bibfnamefont{M.~V.}\ \bibnamefont{Stoitsov}}, \bibinfo
   {author} {\bibfnamefont{J.}~\bibnamefont{Dobaczewski}}, \bibinfo {author}
   {\bibfnamefont{W.}~\bibnamefont{Nazarewicz}},\ and\ \bibinfo {author}
   {\bibfnamefont{P.}~\bibnamefont{Ring}},\ }%
   \bibfield{journal}{%
   \Doi{http://dx.doi.org/10.1016/j.cpc.2005.01.001}{\bibinfo {journal} {Comp.\
   Phys.\ Commun.}}\ }%
   \textbf{\bibinfo {volume} {167}},\ \bibinfo {pages} {43} (\bibinfo {year}
   {2005})%
   \bibAnnoteFile{NoStop}{Stoitsov2005}%
 \bibitem{Delion1996}%
   \BibitemOpen
   \bibfield{author}{%
   \bibinfo {author} {\bibfnamefont{D.~S.}\ \bibnamefont{Delion}}, \bibinfo
   {author} {\bibfnamefont{A.}~\bibnamefont{Insolia}},\ and\ \bibinfo {author}
   {\bibfnamefont{R.~J.}\ \bibnamefont{Liotta}},\ }%
   \bibfield{journal}{%
   \Doi{10.1103/PhysRevC.54.292}{\bibinfo {journal} {Phys.\ Rev.\ C}}\ }%
   \textbf{\bibinfo {volume} {54}},\ \bibinfo {pages} {292} (\bibinfo {year}
   {1996})%
   \bibAnnoteFile{NoStop}{Delion1996}%
 \bibitem{Delion2000}%
   \BibitemOpen
   \bibfield{author}{%
   \bibinfo {author} {\bibfnamefont{D.~S.}\ \bibnamefont{Delion}}\ and\ \bibinfo
   {author} {\bibfnamefont{J.}~\bibnamefont{Suhonen}},\ }%
   \bibfield{journal}{%
   \Doi{10.1103/PhysRevC.61.024304}{\bibinfo {journal} {Phys. Rev. C}}\ }%
   \textbf{\bibinfo {volume} {61}},\ \bibinfo {pages} {024304} (\bibinfo {year}
   {2000})%
   \bibAnnoteFile{NoStop}{Delion2000}%
 \bibitem{Insolia1988}%
   \BibitemOpen
   \bibfield{author}{%
   \bibinfo {author} {\bibfnamefont{A.}~\bibnamefont{Insolia}}, \bibinfo
   {author} {\bibfnamefont{R.~J.}\ \bibnamefont{Liotta}},\ and\ \bibinfo
   {author} {\bibfnamefont{E.}~\bibnamefont{Maglione}},\ }%
   \bibfield{journal}{%
   \Doi{10.1209/0295-5075/7/3/004}{\bibinfo {journal} {Europhys. Lett.}}\ }%
   \textbf{\bibinfo {volume} {7}},\ \bibinfo {pages} {209} (\bibinfo {year}
   {1988})%
   \bibAnnoteFile{NoStop}{Insolia1988}%
 \bibitem{AbramowitzStegun}%
   \BibitemOpen
   \bibfield{author}{%
   \bibinfo {author} {\bibfnamefont{M.}~\bibnamefont{Abramowitz}}\ and\ \bibinfo
   {author} {\bibfnamefont{I.}~\bibnamefont{Stegun}},\ }%
   \emph{\bibinfo {title} {Handbook of Mathematical Functions}},\ Applied
   mathematics series\ (\bibinfo {publisher} {National Bureau of Standards},\
   \bibinfo {year} {1972})%
   \bibAnnoteFile{NoStop}{AbramowitzStegun}%
 \bibitem{Gamow1928}%
   \BibitemOpen
   \bibfield{author}{%
   \bibinfo {author} {\bibfnamefont{G.}~\bibnamefont{Gamow}},\ }%
   \bibfield{journal}{%
   \Doi{10.1007/BF01343196}{\bibinfo {journal} {Z. Phys.}}\ }%
   \textbf{\bibinfo {volume} {51}},\ \bibinfo {pages} {204} (\bibinfo {year}
   {1928})%
   \bibAnnoteFile{NoStop}{Gamow1928}%
 \bibitem{HumbletRosenfeld1961}%
   \BibitemOpen
   \bibfield{author}{%
   \bibinfo {author} {\bibfnamefont{J.}~\bibnamefont{Humblet}}\ and\ \bibinfo
   {author} {\bibfnamefont{L.}~\bibnamefont{Rosenfeld}},\ }%
   \bibfield{journal}{%
   \Doi{http://dx.doi.org/10.1016/0029-5582(61)90207-3}{\bibinfo {journal}
   {Nuclear Physics}}\ }%
   \textbf{\bibinfo {volume} {26}},\ \bibinfo {pages} {529 } (\bibinfo {year}
   {1961})%
   \bibAnnoteFile{NoStop}{HumbletRosenfeld1961}%
 \bibitem{Thomas1954}%
   \BibitemOpen
   \bibfield{author}{%
   \bibinfo {author} {\bibfnamefont{R.~G.}\ \bibnamefont{Thomas}},\ }%
   \bibfield{journal}{%
   \Doi{10.1143/PTP.12.253}{\bibinfo {journal} {Prog.\ of theo.\ phys.}}\ }%
   \textbf{\bibinfo {volume} {12}},\ \bibinfo {pages} {253} (\bibinfo {year}
   {1954})%
   \bibAnnoteFile{NoStop}{Thomas1954}%
 \bibitem{KruppaNazarewicz2004}%
   \BibitemOpen
   \bibfield{author}{%
   \bibinfo {author} {\bibfnamefont{A.~T.}\ \bibnamefont{Kruppa}}\ and\ \bibinfo
   {author} {\bibfnamefont{W.}~\bibnamefont{Nazarewicz}},\ }%
   \bibfield{journal}{%
   \Doi{10.1103/PhysRevC.69.054311}{\bibinfo {journal} {Phys. Rev. C}}\ }%
   \textbf{\bibinfo {volume} {69}},\ \bibinfo {pages} {054311} (\bibinfo {year}
   {2004})%
   \bibAnnoteFile{NoStop}{KruppaNazarewicz2004}%
 \bibitem{Zeh1963}%
   \BibitemOpen
   \bibfield{author}{%
   \bibinfo {author} {\bibfnamefont{H.-D.}\ \bibnamefont{Zeh}},\ }%
   \bibfield{journal}{%
   \Doi{10.1007/BF01375342}{\bibinfo {journal} {Z. Phys.}}\ }%
   \textbf{\bibinfo {volume} {175}},\ \bibinfo {pages} {490} (\bibinfo {year}
   {1963})%
   \bibAnnoteFile{NoStop}{Zeh1963}%
 \bibitem{Mang1964}%
   \BibitemOpen
   \bibfield{author}{%
   \bibinfo {author} {\bibfnamefont{H.~J.}\ \bibnamefont{Mang}},\ }%
   \bibfield{journal}{%
   \Doi{10.1146/annurev.ns.14.120164.000245}{\bibinfo {journal} {Annu.\ Rev.\
   Nucl.\ Sci.}}\ }%
   \textbf{\bibinfo {volume} {14}},\ \bibinfo {pages} {1} (\bibinfo {year}
   {1964})%
   \bibAnnoteFile{NoStop}{Mang1964}%
 \bibitem{Bohm1989}%
   \BibitemOpen
   \bibfield{author}{%
   \bibinfo {author} {\bibfnamefont{A.}~\bibnamefont{Bohm}}, \bibinfo {author}
   {\bibfnamefont{M.}~\bibnamefont{Gadella}},\ and\ \bibinfo {author}
   {\bibfnamefont{G.~B.}\ \bibnamefont{Mainland}},\ }%
   \bibfield{journal}{%
   \Doi{10.1119/1.15797}{\bibinfo {journal} {Am. J. Phys.}}\ }%
   \textbf{\bibinfo {volume} {57}},\ \bibinfo {pages} {1103} (\bibinfo {year}
   {1989})%
   \bibAnnoteFile{NoStop}{Bohm1989}%
 \bibitem{Tonozuka1979}%
   \BibitemOpen
   \bibfield{author}{%
   \bibinfo {author} {\bibfnamefont{I.}~\bibnamefont{Tonozuka}}\ and\ \bibinfo
   {author} {\bibfnamefont{A.}~\bibnamefont{Arima}},\ }%
   \bibfield{journal}{%
   \Doi{http://dx.doi.org/10.1016/0375-9474(79)90415-9}{\bibinfo {journal}
   {Nucl. Phys.}}\ }%
   \textbf{\bibinfo {volume} {A323}},\ \bibinfo {pages} {45 } (\bibinfo {year}
   {1979})%
   \bibAnnoteFile{NoStop}{Tonozuka1979}%
 \bibitem{Eichler1965}%
   \BibitemOpen
   \bibfield{author}{%
   \bibinfo {author} {\bibfnamefont{J.}~\bibnamefont{Eichler}}\ and\ \bibinfo
   {author} {\bibfnamefont{H.-J.}\ \bibnamefont{Mang}},\ }%
   \bibfield{journal}{%
   \Doi{10.1007/BF01393265}{\bibinfo {journal} {Z. Phys.}}\ }%
   \textbf{\bibinfo {volume} {183}},\ \bibinfo {pages} {321} (\bibinfo {year}
   {1965})%
   \bibAnnoteFile{NoStop}{Eichler1965}%
 \bibitem{Kamuntavicius2001}%
   \BibitemOpen
   \bibfield{author}{%
   \bibinfo {author} {\bibfnamefont{G.}~\bibnamefont{Kamuntavi{\v{c}}ius}},
   \bibinfo {author} {\bibfnamefont{R.}~\bibnamefont{Kalinauskas}}, \bibinfo
   {author} {\bibfnamefont{B.}~\bibnamefont{Barret}}, \bibinfo {author}
   {\bibfnamefont{S.}~\bibnamefont{Mickevi{\v{c}}ius}},\ and\ \bibinfo {author}
   {\bibfnamefont{D.}~\bibnamefont{Germanas}},\ }%
   \bibfield{journal}{%
   \Doi{10.1016/S0375-9474(01)01101-0}{\bibinfo {journal} {Nucl.\ Phys.}}\ }%
   \textbf{\bibinfo {volume} {A695}},\ \bibinfo {pages} {191 } (\bibinfo {year}
   {2001})%
   \bibAnnoteFile{NoStop}{Kamuntavicius2001}%
 \bibitem{VautherinBrink1972}%
   \BibitemOpen
   \bibfield{author}{%
   \bibinfo {author} {\bibfnamefont{D.}~\bibnamefont{Vautherin}}\ and\ \bibinfo
   {author} {\bibfnamefont{D.~M.}\ \bibnamefont{Brink}},\ }%
   \bibfield{journal}{%
   \Doi{10.1103/PhysRevC.5.626}{\bibinfo {journal} {Phys. Rev. C}}\ }%
   \textbf{\bibinfo {volume} {5}},\ \bibinfo {pages} {626} (\bibinfo {year}
   {1972})%
   \bibAnnoteFile{NoStop}{VautherinBrink1972}%
 \bibitem{Bennaceur2005}%
   \BibitemOpen
   \bibfield{author}{%
   \bibinfo {author} {\bibfnamefont{K.}~\bibnamefont{Bennaceur}}\ and\ \bibinfo
   {author} {\bibfnamefont{J.}~\bibnamefont{Dobaczewski}},\ }%
   \bibfield{journal}{%
   \Doi{10.1016/j.cpc.2005.02.002}{\bibinfo {journal} {Comp. Phys. Commun.}}\ }%
   \textbf{\bibinfo {volume} {168}},\ \bibinfo {pages} {96 } (\bibinfo {year}
   {2005})%
   \bibAnnoteFile{NoStop}{Bennaceur2005}%
 \bibitem{Akovali1998}%
   \BibitemOpen
   \bibfield{author}{%
   \bibinfo {author} {\bibfnamefont{Y.}~\bibnamefont{Akovali}},\ }%
   \bibfield{journal}{%
   \Doi{10.1006/ndsh.1998.0009}{\bibinfo {journal} {Nuclear Data Sheets}}\ }%
   \textbf{\bibinfo {volume} {84}},\ \bibinfo {pages} {1 } (\bibinfo {year}
   {1998})%
   \bibAnnoteFile{NoStop}{Akovali1998}%
 \bibitem{Samyn2003}%
   \BibitemOpen
   \bibfield{author}{%
   \bibinfo {author} {\bibfnamefont{M.}~\bibnamefont{Samyn}}, \bibinfo {author}
   {\bibfnamefont{S.}~\bibnamefont{Goriely}},\ and\ \bibinfo {author}
   {\bibfnamefont{J.}~\bibnamefont{Pearson}},\ }%
   \bibfield{journal}{%
   \Doi{http://dx.doi.org/10.1016/S0375-9474(03)01578-1}{\bibinfo {journal}
   {Nucl. Phys.}}\ }%
   \textbf{\bibinfo {volume} {A725}},\ \bibinfo {pages} {69 } (\bibinfo {year}
   {2003})%
   \bibAnnoteFile{NoStop}{Samyn2003}%
 \bibitem{Chabanat1998}%
   \BibitemOpen
   \bibfield{author}{%
   \bibinfo {author} {\bibfnamefont{E.}~\bibnamefont{Chabanat}}, \bibinfo
   {author} {\bibfnamefont{P.}~\bibnamefont{Bonche}}, \bibinfo {author}
   {\bibfnamefont{P.}~\bibnamefont{Haensel}}, \bibinfo {author}
   {\bibfnamefont{J.}~\bibnamefont{Meyer}},\ and\ \bibinfo {author}
   {\bibfnamefont{R.}~\bibnamefont{Schaeffer}},\ }%
   \bibfield{journal}{%
   \Doi{10.1016/S0375-9474(98)00180-8}{\bibinfo {journal} {Nucl.\ Phys.}}\ }%
   \textbf{\bibinfo {volume} {A635}},\ \bibinfo {pages} {231; 643, 441(E)}
   (\bibinfo {year} {1998})%
   \bibAnnoteFile{NoStop}{Chabanat1998}%
 \bibitem{Moller1995}%
   \BibitemOpen
   \bibfield{author}{%
   \bibinfo {author} {\bibfnamefont{P.}~\bibnamefont{Moller}}, \bibinfo {author}
   {\bibfnamefont{J.}~\bibnamefont{Nix}}, \bibinfo {author}
   {\bibfnamefont{W.}~\bibnamefont{Myers}},\ and\ \bibinfo {author}
   {\bibfnamefont{W.}~\bibnamefont{Swiatecki}},\ }%
   \bibfield{journal}{%
   \Doi{10.1006/adnd.1995.1002}{\bibinfo {journal} {At. Data Nucl. Data
   Tables}}\ }%
   \textbf{\bibinfo {volume} {59}},\ \bibinfo {pages} {185 } (\bibinfo {year}
   {1995})%
   \bibAnnoteFile{NoStop}{Moller1995}%
 \bibitem{Qi2009}%
   \BibitemOpen
   \bibfield{author}{%
   \bibinfo {author} {\bibfnamefont{C.}~\bibnamefont{Qi}}, \bibinfo {author}
   {\bibfnamefont{F.~R.}\ \bibnamefont{Xu}}, \bibinfo {author}
   {\bibfnamefont{R.~J.}\ \bibnamefont{Liotta}}, \bibinfo {author}
   {\bibfnamefont{R.}~\bibnamefont{Wyss}}, \bibinfo {author}
   {\bibfnamefont{M.~Y.}\ \bibnamefont{Zhang}}, \bibinfo {author}
   {\bibfnamefont{C.}~\bibnamefont{Asawatangtrakuldee}},\ and\ \bibinfo {author}
   {\bibfnamefont{D.}~\bibnamefont{Hu}},\ }%
   \bibfield{journal}{%
   \Doi{10.1103/PhysRevC.80.044326}{\bibinfo {journal} {Phys. Rev. C}}\ }%
   \textbf{\bibinfo {volume} {80}},\ \bibinfo {pages} {044326} (\bibinfo {year}
   {2009})%
   \bibAnnoteFile{NoStop}{Qi2009}%
 \bibitem{MuntianPatykSobiczewski2003}%
   \BibitemOpen
   \bibfield{author}{%
   \bibinfo {author} {\bibfnamefont{I.}~\bibnamefont{Muntian}}, \bibinfo
   {author} {\bibfnamefont{Z.}~\bibnamefont{Patyk}},\ and\ \bibinfo {author}
   {\bibfnamefont{A.}~\bibnamefont{Sobiczewski}},\ }%
   \bibfield{journal}{%
   \Doi{10.1134/1.1586412}{\bibinfo {journal} {Phys. Atom. Nucl.}}\ }%
   \textbf{\bibinfo {volume} {66}},\ \bibinfo {pages} {1015} (\bibinfo {year}
   {2003})%
   \bibAnnoteFile{NoStop}{MuntianPatykSobiczewski2003}%
 \bibitem{SobiczewskiPrivateComm}%
   \BibitemOpen
   \bibfield{author}{%
   \bibinfo {author} {\bibfnamefont{A.}~\bibnamefont{Sobiczewski}},\ }%
   \bibinfo {howpublished} {Private communication}%
   \bibAnnoteFile{NoStop}{SobiczewskiPrivateComm}%
 \bibitem{DelionSandulescu2002}%
   \BibitemOpen
   \bibfield{author}{%
   \bibinfo {author} {\bibfnamefont{D.~S.}\ \bibnamefont{Delion}}\ and\ \bibinfo
   {author} {\bibfnamefont{A.}~\bibnamefont{Sandulescu}},\ }%
   \bibfield{journal}{%
   \Doi{10.1088/0954-3899/28/4/303}{\bibinfo {journal} {J. Phys G}}\ }%
   \textbf{\bibinfo {volume} {28}},\ \bibinfo {pages} {617} (\bibinfo {year}
   {2002})%
   \bibAnnoteFile{NoStop}{DelionSandulescu2002}%
 \bibitem{Betan2012}%
   \BibitemOpen
   \bibfield{author}{%
   \bibinfo {author} {\bibfnamefont{R.~I.}\ \bibnamefont{Betan}}\ and\ \bibinfo
   {author} {\bibfnamefont{W.}~\bibnamefont{Nazarewicz}},\ }%
   \bibfield{journal}{%
   \Doi{10.1103/PhysRevC.86.034338}{\bibinfo {journal} {Phys. Rev. C}}\ }%
   \textbf{\bibinfo {volume} {86}},\ \bibinfo {pages} {034338} (\bibinfo {year}
   {2012})%
   \bibAnnoteFile{NoStop}{Betan2012}%
 \bibitem{Fliessbach1975}%
   \BibitemOpen
   \bibfield{author}{%
   \bibinfo {author} {\bibfnamefont{T.}~\bibnamefont{Fliessbach}},\ }%
   \bibfield{journal}{%
   \Doi{10.1007/BF01408426}{\bibinfo {journal} {Z. Phys. A}}\ }%
   \textbf{\bibinfo {volume} {272}},\ \bibinfo {pages} {39} (\bibinfo {year}
   {1975})%
   \bibAnnoteFile{NoStop}{Fliessbach1975}%
 \bibitem{FliessbachMang1976}%
   \BibitemOpen
   \bibfield{author}{%
   \bibinfo {author} {\bibfnamefont{T.}~\bibnamefont{Fliessbach}}\ and\ \bibinfo
   {author} {\bibfnamefont{H.}~\bibnamefont{Mang}},\ }%
   \bibfield{journal}{%
   \Doi{http://dx.doi.org/10.1016/0375-9474(76)90184-6}{\bibinfo {journal}
   {Nucl. Phys.}}\ }%
   \textbf{\bibinfo {volume} {A263}},\ \bibinfo {pages} {75 } (\bibinfo {year}
   {1976})%
   \bibAnnoteFile{NoStop}{FliessbachMang1976}%
 \bibitem{TonozukaArima1979}%
   \BibitemOpen
   \bibfield{author}{%
   \bibinfo {author} {\bibfnamefont{I.}~\bibnamefont{Tonozuka}}\ and\ \bibinfo
   {author} {\bibfnamefont{A.}~\bibnamefont{Arima}},\ }%
   \bibfield{journal}{%
   \Doi{10.1016/0375-9474(79)90415-9}{\bibinfo {journal} {Nucl. Phys.}}\ }%
   \textbf{\bibinfo {volume} {A323}},\ \bibinfo {pages} {45 } (\bibinfo {year}
   {1979})%
   \bibAnnoteFile{NoStop}{TonozukaArima1979}%
 \bibitem{Giuliani2002}%
   \BibitemOpen
   \bibfield{author}{%
   \bibinfo {author} {\bibfnamefont{G.~F.}\ \bibnamefont{Giuliani}}\ and\
   \bibinfo {author} {\bibfnamefont{G.}~\bibnamefont{Vignale}},\ }%
   \emph{\bibinfo {title} {Quantum theory of the electron liquid}}\ (\bibinfo
   {publisher} {Cambridge university press, Cambridge},\ \bibinfo {year}
   {2005})%
   \bibAnnoteFile{NoStop}{Giuliani2002}%
 \bibitem{RingSchuck1980}%
   \BibitemOpen
   \bibfield{author}{%
   \bibinfo {author} {\bibfnamefont{P.}~\bibnamefont{Ring}}\ and\ \bibinfo
   {author} {\bibfnamefont{P.}~\bibnamefont{Schuck}},\ }%
   \emph{\bibinfo {title} {The nuclear many body problem}}\ (\bibinfo
   {publisher} {1st ed. Springer-Verlag, New York},\ \bibinfo {year} {1980})%
   \bibAnnoteFile{NoStop}{RingSchuck1980}%
 \end{thebibliography}

 %

\end{document}